\documentclass[useAMS,usenatbib,aastex]{mnras}

\usepackage{graphicx,rotating}
\usepackage{amssymb,amsmath,pdflscape}
\usepackage{natbib}

\newcommand{\tf}{t_\text{final}}

\newcommand{\hu}{\hat{U}}
\newcommand{\hv}{\hat{V}}
\newcommand{\hw}{\hat{W}}
\newcommand{\hcin}{\hat{T}}
\newcommand{\hpi}{\hat{\Pi}}

\newcommand{\hjcin}{\hat{J}}
\newcommand{\hpsi}{\hat{\Psi}}
\newcommand{\hphi}{\hat{\Phi}}
\newcommand{\hper}{\hat{p}}

\newcommand{\hp}{\hat{P}}

\newcommand{\hrho}{\hat{\rho}}

\newcommand{\hvol}{\hat{V}}
\newcommand{\hm}{\hat{M}}

\newcommand{\hh}{\hat{H}}
\newcommand{\hnorm}{\hat{H}_{\rm norm.}}
\newcommand{\hr}{\hat{R}}
\newcommand{\hd}{\hat{d}}

\newcommand{\hz}{\hat{z}}

\newcommand{\hrs}{\hat{r}}
\newcommand{\ho}{\hat{\Omega}}

\newcommand{\hba}{\hat{A}}
\newcommand{\hbsur}{\hat{S}}

\newcommand{\pmax}{\hp_{\rm max}}
\newcommand{\rhomax}{\hrho_{\rm max}}
\newcommand{\hmax}{\hh_{\rm max}}

\begin{document}

\title{The equilibrium of over-pressurised polytropes}

\author[J.-M. Hur\'e, F. Hersant and G. Nasello]
{J.-M. Hur\'e$^{1,2}$\thanks{E-mail:jean-marc.hure@u-bordeaux.fr},
F. Hersant$^{1,2}$,
and G. Nasello$^{3}$\\
$^{1}$Univ. Bordeaux, LAB, UMR 5804, F-33615, Pessac, France\\
$^{2}$CNRS, LAB, UMR 5804, F-33615, Pessac, France\\
$^{3}$Institut Utinam, CNRS UMR 6213, Universit\'e de Franche-Comt\'e, OSU THETA Franche-Comt\'e-Bourgogne,\\ Observatoire de Besan\c{c}on, BP 1615, 25010 Besan\c{c}on Cedex, France}

\date{Received ??? / Accepted ???}
 
\pagerange{\pageref{firstpage}--\pageref{lastpage}} \pubyear{???}

\maketitle

\label{firstpage}

\begin{abstract}
We investigate the impact of an external pressure on the structure of self-gravitating polytropes for axially symmetric ellipsoids and rings. The confinement of the fluid by photons is accounted for through a boundary condition on the enthalpy $H$. Equilibrium configurations are determined numerically from a generalised ``Self-Consistent-Field''-method. The new algorithm incorporates an intra-loop re-scaling operator ${\cal R}(H)$, which is essential for both convergence and getting self-normalised solutions. The main control parameter is the external-to-core enthalpy ratio. In the case of uniform rotation rate and uniform surrounding pressure, we compute the mass, the volume, the rotation rate and the maximum enthalpy. This is repeated for a few polytropic indices $n$. For a given axis ratio, over-pressurization globally increases all output quantities, and this is more pronounced for large $n$. Density profiles are flatter than in the absence of an external pressure. When the control parameter asymptotically tends to unity, the fluid converges toward the incompressible solution, whatever the index, but becomes geometrically singular. Equilibrium sequences, obtained by varying the axis ratio, are built. States of critical rotation are greatly exceeded or even disappear. The same trends are observed with differential rotation. Finally, the typical response to a photon point source is presented. Strong irradiation favours sharp edges. Applications concern star forming regions and matter orbiting young stars and black holes.
\end{abstract}


\section{Introduction}

Light travelling across the Universe is an unavoidable source of energy that perturbs its constituents. The close neighbourhood of stars and compact objects is clearly a privileged place where the intense radiation hits, penetrates and heats up the surrounding companion stars, accretion disks, and planets and the interstellar medium as well \citep[e.g.][]{smak89,tout89,seager98,rozanska02}. Over-pressurization by gas is recognised as an important mechanism capable of shaping giant molecular clouds and triggering star formation \citep{um99,ik00,bo10,maruta10,kaminski14,bieri16}. Pressure by photons is an obstacle for massive proto-stars to accreting large amounts of gas, and for massive and super-massive stars to evolving steadily \citep{langer97,dotan12}. It prevents the fast growth of super-massive black holes in Active Galactic Nuclei \citep{ka04,coza08}.

From a theoretical point of view, the problem of over-pressurization can take various forms. \cite{milne23,milne36a,milne36b} soon investigated the mechanical connection between the interior of a static star and its upper photosphere. \cite{ebert55}, \cite{bonnor56} and \cite{mcc57} have calculated the stability of an isothermal sphere subject to external pressure in terms of critical pressure, radius and mass. This kind of analysis has been extended to gas obeying a polytropic equation-of-state in \cite{ho70} and \cite{wa81}, while \cite{sipila11} have considered an ideal gas. According to \cite{ui86}, background matter has a stabilising effect for polytropic indices $n \ge 3$.

Complexity is increased by rotation in several respects. This is, first of all, a technical difficulty. The problem becomes at least bi-dimensional. With a non-spherical fluid boundary, the gravity field can no longer be determined from the Gauss theorem.  Second, external matter that exerts pressure on the system also participates in the force budget. This point is generally ignored, leaving a certain inconsistency \citep{horedttextbook2004}. Finally, the rotation law remains more or less ad-hoc \citep[e.g.][]{sta83a,hachisu86,th90}. \cite{weber76} has investigated the impact of external pressure on rigidly rotating, homogeneous fluids from the tensor Virial equations \citep{chandra73,cl62}; see also \cite{mpt77}. He pointed out the importance of the total angular momentum and symmetries (axial or tri-axial) on equilibrium states and their compressional stability. In \cite{viala78}, stability is studied for $n<0$, and the influence of rotation is shown to be weak unless the gas mass exceeds several hundreds solar masses. The structure of rotating isothermal clouds has been computed in \cite{sta83a,sta83b} for a series of masses and kinetic-to-gravitational energy ratios, and the stability is discussed mainly through total mass-mass density diagrams.

In this article, we compute the structure of a rotating polytropic fluid undergoing a mechanical pressure $P_e$ due to photons. Without central symmetry, matter outside the fluid creates an additional potential that influences the fluid structure \citep[e.g.][]{ui86}. This difficulty does not exist with mass-less particles : that is why we consider that photons are the source of confinement. We assume that the fluid is perfectly reflecting, i.e. photons do not penetrate below the fluid boundary and do not bring extra energy to the system. A two-layer model would be more realistic \citep[e.g.][]{ca86,ru88,cmc00,kong10,kiu10}. No stability analysis will be presented here \citep[e.g.][]{th90, cen01, sy16}. A particular motivation for this work is to determine the impact of $P_e$ in the classical angular momentum-rotation rate diagram \citep{hachisu86}. Except in the incompressible case, equilibrium sequences are open with end-points beyond which no physical state exists. To what extent is external pressure capable of modifying this picture ? Another interesting aspect concerns the response of the fluid when the ambient pressure is not uniform along the fluid boundary. This is a common situation in the universe, e.g. gas and dust cloud in the vicinity of a bright star, a disc orbiting a young star, the environment of a black hole, etc. Finally, solving the problem in the framework of the ``Self-Consistent-Field''-method is an interesting technical challenge that, to our knowledge, has never been adressed; see \cite{ca16} for a full, multi-component analytical approach.

In practice, the problem can be cast in a fully differential form (i.e., a two-dimensional version of the Lane-Emden equation) or in a more algebraic form, namely the Bernoulli equation coupled to the Poisson equation, which is the option selected here. These two approaches, however, do not always share the same families of solutions. We recall the basic equation set in Sect. \ref{sec:overp}, and show that the pressure balance at the fluid boundary is equivalent to an enthalpy balance, due to the polytropic assumption. The formulae for the main constants of the problem are derived. Expected deviations with respect to the zero-pressure case are discussed. As argued in Sect. \ref{sec:gscf}, equilibrium states can be numerically captured from the ``Self-Consistent-Field''-method \citep{bo73,hachisu86}, but the standard version is not operational and must be extended. We propose a more general algorithm that works for any spatially-dependent external stress. In particular, it is well suited to situations where the fluid is located next to a photon source, which is a situation of great astrophysical interest. The details of the numerical procedure are briefly outlined in Sect. \ref{sec:ans}. Section \ref{sec:uniformp} is devoted to the results obtained assuming rigid rotation and uniform radiation pressure. We report the relationships for the fluid volume, mass and rotation rate vs. the external-to-central pressure ratio. We discuss both ellipsoidal and toroidal configurations for various indices. In Sect. \ref{eq:eseq}, we determine the new equilibrium sequences for $n=0.5$ and $n=1.5$ by varying the fluid axis-ratio while holding $P_e$ fixed, and vice-versa. The classical rotation rate-angular momentum diagram is strongly impacted, especially the mass-shedding limits. The kinetic-to-gravitational energy ratio, which is the classical indicator for dynamical stabilities, is deduced. We consider differential rotation in Sect. \ref{sec:difrot}, namely the classical $v$- and $j$-constant profiles. We finally illustrate the capacity of the method by considering a fluid (ellipsoid and ring) illuminated by a point source located on the rotation axis. This is the aim of Sect. \ref{sec:ips}. A few concluding remarks and perspectives are given in the last section.

\section{The over-pressurized fluid}
\label{sec:overp}

\subsection{The dimensionless equation set}

\begin{figure}
\includegraphics[width=8.4cm,bb=0 0 475 331,clip==]{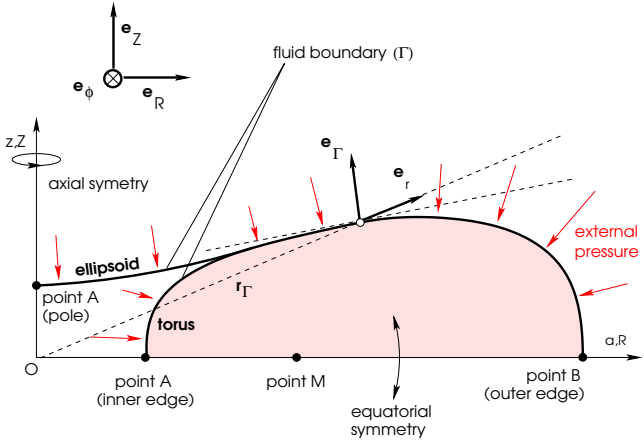}
\caption{Configuration for the self-gravitating fluid (ellipsoid or torus), limited to the upper plane $Z>0$. Axial and equatorial symmetries are assumed. Photons exert a mechanical pressure $P_e$ along the fluid boundary $(\Gamma)$ ({\it red arrows}). Reference points A, B and M used in the SCF-method are indicated.}
\label{fig:tore_et_ell.eps}
\end{figure}

The framework, hypothesis and notations are the same as in \cite{hh17} (hereafter, Paper I): the fluid is a self-gravitating polytrope rotating steadily in an imposed centrifugal potential (or rotation law). We assume both equatorial and axial symmetries and focus on single body equilibria. Cylindrical coordinates $(R,Z)$ are used. The system is depicted in Fig. \ref{fig:tore_et_ell.eps}. Gas pressure $P$ and mass density $\rho$ are linked through a polytropic equation of state $P=K\rho^\gamma$, where $K$ and $\gamma$ are positive constants. The rotation rate $\Omega(R)$ depends solely on the radial coordinate $R$, due to the integrability condition \citep{amendt1989}. The relevant equation set is composed of : i) a space invariant involving the enthalpy $H=\int{\frac{d P}{\rho}}$, the gravitational potential $\Psi$ and the centrifugal potential $\Phi$ \citep[e.g][]{bo73,em85,hachisu86}, ii) the enthalpy-density relationship which follows from the polytropic assumption, and iii) the Poisson equation linking the mass distribution to the gravitational potential. In dimensionless form, this set is
\begin{flalign}
\begin{cases}
\hpsi + C_1\hh + C_2 \hphi = C_3,\\
\hrho^{1/n}=\sup(\hh,0),\\
\Delta \hpsi = 4 \pi \hrho,\\
\end{cases}
\label{eq:eset}
\end{flalign}
where $n=\frac{1}{\gamma-1}$ is the polytropic index (we focus on positive indices). The mass density of the fluid is associated with positive values of the enthalpy field, which is the role the supremum function in Eq.(\ref{eq:eset}b).

The two constants $C_1$ and $C_2$ contain the three physical scales, namely the length $L$ of the system, the mass $\rho_0 L^3$ and the orbital time scale $1/\Omega_0$. We have
\begin{flalign}
\begin{cases}
C_1\equiv \frac{ K (n+1)\rho_0^{\gamma-2}}{GL^2},\\
C_2\equiv \frac{\Omega^2_0}{G\rho_0},\\
\end{cases}
\label{eq:c1c2c3}
\end{flalign}
for finite indices. Going back to a particular physical problem requires the specification of only two quantities, for instance $L$ and $\rho_0$ (and $\Omega_0$ follows). The scale-free approach is very powerful as it unifies all possible solutions. From a numerical point of view, adimensioning enables to work with quantities and fields of the order of unity. In particular, $\hh \in [0,1]$ is assumed, and so on for $\hrho$ according to Eq.(\ref{eq:eset}b).

\subsection{Hypothesis on the source of confinement}

Under central symmetry (i.e. without any rotation), matter located beyond the fluid boundary does not influence the internal gravitational potential and forces. This is a consequence of the Gauss theorem \citep[e.g.][]{kellogg29}. So, when the surrounding medium exerts some pressure, the volume occupied by the fluid is limited by the spherical shell where the internal pressure meets the external pressure $P_e$, namely
\begin{equation}
P(\Gamma)=P_e(\vec{r}_\Gamma),
\label{eq:pbal}
\end{equation}
where $\vec{r}_\Gamma$ refers to the location of the boundary ($\Gamma$).

For rotating systems, central symmetry is broken. External matter exerts non-trivial forces that do not cancel inside the fluid and can not be ignored. This difficulty is bypassed if the source of external pressure is mass-less, i.e. made of photons. This is the assumption made here. In reality, photons penetrate more or less deeply inside matter depending on its absorption capabilities, thereby creating a photosphere. At the lowest order, this means a double-layer fluid, i.e. a core where gas pressure dominates surrounded by a radiation-pressure dominated envelope. Multi-polytropic models are commonly constructed for stars, planets and discs and offer better realism \citep{milne36b,bee88,ru88,cmc00,du03,kong10,kiu10,remus15,ka16}. This approach would be more realistic, but it is out of the scope of the present study. For simplicity, we consider that photons deposit momentum at the surface and are totally reflected (albedo is unity). There is therefore no energy accumulation at the fluid boundary.

\subsection{Pressure balance at the fluid boundary}
\label{subsec.pbatfb}

At equilibrium, pressure balance must be fulfilled at any point of the boundary, be the fluid isolated or not. This condition must be included in the above equation set. We see from Eq.(\ref{eq:eset}b) that an external enthalpy $H_e$ is naturally associated to any value of $P_e$, i.e.
\begin{equation}
\hh_e^{1+n} = \hp_e,
\label{eq:ph}
\end{equation}
and so Eq.(\ref{eq:pbal}) becomes
\begin{equation}
\hh(\Gamma) - \hh_e(\vec{r}_\Gamma) = 0.
\label{eq:hbal}
\end{equation}
It follows that the mass density below the fluid surface corresponds to the enthalpy level above $\hh_e$, namely
\begin{equation}
\hrho^\frac{1}{n} = \sup \left(\hh,\hh_e\right) > 0,
\label{eq:rhoh_sup_pe}
\end{equation}
which replaces Eq.(\ref{eq:eset}b). The mass density is now positive onto $(\Gamma)$ as soon as $\hh_e >0$. It is therefore locally discontinuous, like for the isolated case with $n=0$ (e.g. Maclaurin or Jacobi ellipsoids). This is not a new result. The situation is similar when the fluid is embedded in a ambient medium \citep{ui86,ho00,km16}, but there is no background matter here (over-pressure is due to photons). Note that $\hh_e$ can vary in space. This is for instance the case of a molecular cloud illuminated by a nearby star, a circumstellar disk or binaries. Certain parts of the fluid receive photons, while other parts do not (see Sect. \ref{sec:ips}).

External pressure (and subsequently $\hh_e$) imposed by the environment is, in principle, unlimited. However, $\hh_e$ must be less that the maximal value $\hh_{\rm max}=\max (\hh)$ of the enthalpy inside the fluid, otherwise the fluid no more exists. In the limit $\hh_e \rightarrow \hh_{\rm max}$, there are a priori two possible configurations, independent of the polytropic index. Either, the fluid is reduced to an infinitely thin medium because ablation of all fluid layers leaves only one point of the core where Eq.(\ref{eq:hbal}) is satisfied \citep{petroff08}. In this case, the potential diverges, and so does $\hh$. Or $\hh = \hh_{\rm max} = \hh_e$ everywhere inside the entire fluid that becomes fully homogeneous (like in the isolated case with $n=0$). There is, however, a contradiction is this second option. As Eq.(\ref{eq:eset}a) shows, the enthalpy field is the image of the gravitational potential (deformed by $\hphi$). For homogeneous systems, $\Psi$ has a marked well which is not compatible with a flat enthalpy field. The only way to save this picture is to shrink the length scale $L$ to zero. We conclude that, at maximal external pressure, the fluid tends to a singularity (a point or an infinitely thin loop), whatever the polytropic index.  

\subsection{The three constants}
\label{eq:constants}

The general equation set to be solved is
\begin{flalign}
\begin{cases}
\hpsi + C_1\hh + C_2 \hphi = C_3,\\
\hrho^\frac{1}{n} = \sup \left(\hh,\hh_e\right) > 0,\\
\Delta \hpsi = 4 \pi \hrho,\\
\hh(\vec{r}_\Gamma)-\hh_e(\vec{r}_\Gamma)=0,
\end{cases}
\label{eq:eset2}
\end{flalign}
where the last equation stands for the prescription for the external pressure onto $(\Gamma)$. In general, there is very little chance in finding a solution $\hh(\hr,\hz)$ by imposing the three constants $C_1$, $C_2$ and $C_3$ (see paper I).  Actually, Eq.(\ref{eq:eset2}a) gives
\begin{flalign}
\begin{cases}
\hpsi_A + C_1\hh_A + C_2 \hphi_A = C_3,\\
\hpsi_B + C_1\hh_B + C_2 \hphi_B = C_3,\\
\hpsi_M + C_1\hh_M + C_2 \hphi_M = C_3.
\label{eq:bernouilliabc}
\end{cases}
\end{flalign}
Instead, the free parameters are three ``reference'' points A, B and M and associated $\hh$ values (denoted $\hh_A$ and so on). Points A and B are usually located at the fluid boundary while M stands inside where $\hh$ is maximum (see Fig. \ref{fig:tore_et_ell.eps} and below). Solving Eq.(\ref{eq:bernouilliabc}) for the constants leads to
\begin{flalign}
\begin{cases}
C_1=-\frac{\hpsi_M\Delta \hphi_{AB}+\hpsi_A\Delta \hphi_{BM}+\hpsi_B\Delta \hphi_{MA}}{\hh_M\Delta \hphi_{AB}+\hh_A\Delta \hphi_{BM}+\hh_B\Delta \hphi_{MA}},\\\\
C_2=-\frac{\hh_M\Delta \hpsi_{AB}+\hh_A\Delta \hpsi_{BM}+\hh_B\Delta \hpsi_{MA}}{\hh_M\Delta \hphi_{AB}+\hh_A\Delta \hphi_{BM}+\hh_B\Delta \hphi_{MA}},\\\\
C_3= \frac{\hh_B\begin{vmatrix}
\hpsi_A & \hphi_A \\ 
\hpsi_M & \hphi_M\end{vmatrix}+
\hh_M\begin{vmatrix}
\hpsi_B & \hphi_B \\ 
\hpsi_A & \hphi_A\end{vmatrix}+\hh_A\begin{vmatrix}
\hpsi_M & \hphi_M \\ 
\hpsi_B & \hphi_B\end{vmatrix}}{\hh_M \Delta \hphi_{AB}+\hh_A \Delta \hphi_{BM}+\hh_B \Delta \hphi_{MA}}
\end{cases}
\label{eq:ctebc}
\end{flalign}
where $\Delta \hphi_{AB}=\hphi_A-\hphi_B$ and so on for $\Delta \hphi_{BM}$, $\Delta \hphi_{MA}$, $\Delta \hpsi_{AB}$, $\Delta \hpsi_{BM}$, and $\Delta \hpsi_{MA}$. By setting $\hh_A=\hh_B=0$, values for the isolated fluid (superscript '0') are recovered, i.e.
\begin{flalign}
\begin{cases}
C^0_1=-\frac{\hpsi_M\Delta \hphi_{AB}+\hpsi_A\Delta \hphi_{BM}+\hpsi_B\Delta \hphi_{MA}}{\hh_M\Delta \hphi_{AB}},\\
C^0_2=-\frac{\Delta \hpsi_{AB}}{\Delta \hphi_{AB}},\\
C^0_3= \frac{\hpsi_B  \hphi_A - \hphi_B \hpsi_A}{\Delta \hphi_{AB}}.
\end{cases}
\label{eq:ctebc_hahbnull}
\end{flalign}

\subsection{What can we expect ?}
\label{subsec.wcwe}

The consequence of over-pressurization can be anticipated, at least in a qualitative manner. As outlined, the positive mass density on the boundary (or part of it, depending on $\hh_e$) means a flatter profile. So, for a given position of the reference points A and B, and given centrifugal potential $\hphi$, the orbiting dimensionless mass $\hm$ is expected to be larger than in the isolated case. This means a deeper gravitational potential well, and presumably a higher rotation rate, through constant $C_2$. Let us compare two situations : i) the isolated case where $\hh_A=0$ and $\hh_B=0$, and ii) the case of uniform external pressure where $\hh_B-\hh_A=0$ but $\hh_A>0$. From Eqs.(\ref{eq:ctebc}) and (\ref{eq:ctebc_hahbnull}), we get:
\begin{flalign}
\frac{C_2}{C_2^0}=\frac{\Delta \hpsi_{AB}}{\Delta \hpsi_{AB}^0},
\end{flalign}
meaning that $C_2$ changes according to the variation of the potential contrast between the two reference points A and B. For a torus, we are typically dealing with the inner and outer edges. Even, in the limit where A $\rightarrow$ B, these contrasts are nothing but the gravitational accelerations (i.e. the local slopes of $\hpsi$), which are expected to be enhanced as the fluid mass increases. Although it is difficult to be more quantitative, one can reasonably expect $C_2 > C_2^0$. On these grounds, we can infer the behaviour of constant $C_1$ from Eqs.(\ref{eq:bernouilliabc}a) and (\ref{eq:bernouilliabc}b). We have 
\begin{flalign}
\Delta \hpsi_{AM} + C_1\Delta \hh_{AM} + C_2 \Delta \hphi_{AM} = 0.
\end{flalign}
While $\Delta \hh_{AM}=-\hh_M=-1$ in the isolated case (at point M, $\hh$ reaches unity; see below), we have $\Delta \hh_{AM}=\hh_e-1$ with uniform external pressure, and so
\begin{equation}
\frac{C_1}{C_1^0}=\frac{\frac{\Delta \hpsi_{AM}}{\Delta \hphi_{AM}} + C_2}{\frac{\Delta \hpsi^0_{AM}}{\Delta \hphi_{AM}} + C_2^0} \times \frac{1}{1-\hh_e}.
\end{equation}
Because of steeper slopes with external pressure (due to a larger mass), $C_1$ should logically be increased. Even, we have $C_1 \rightarrow \infty$ if $\hh_e \rightarrow 1$ \citep{petroff08}. Finally, regarding constant $C_3$, we find from Eq.(\ref{eq:bernouilliabc}a)
\begin{equation}
C_3-C_3^0=\underbrace{\hpsi_A-\hpsi^0_A+(C_2-C_2^0)\hphi_A}_{\lesssim 0} +\underbrace{C_1 \hh_e}_{\gtrsim 0},
\end{equation}
and the sign depends on the magnitude of the two terms. Predictions are fragile, but if we refer to the limit case where $C_1 \rightarrow \infty$, we can probably have $C_3 \gtrsim C_3^0$. 

To conclude on this simple analysis, over-pressurization is expected to lead to larger values for $C_1$, $C_2$, $|C_3|$ and $\hm$ as well, for a given index $n$. Through Eq.(\ref{eq:c1c2c3}), these trends can be interpreted in terms of the physical parameters for the fluid, i.e. $n$, $K$, $\rho_0$ and $L$. For a given reference density $\rho_0$ for instance, the equilibrium requires a higher rotation rate $\Omega_0$ than in the isolated case, a larger mass (because $\rho$ is larger at the fluid boundary), a larger polytropic constant $K$ or/and a smaller length scale $L$. These should hold whatever the rotation profile $\Omega(R)$.

\section{Generalising the SCF-method}
\label{sec:gscf}

\subsection{Principle and limitation of the standard algorithm}

As soon as $\hh_e(\Gamma)$ is known in advance, there is a priori no obstacle in using the ``Self-Consistent Field'' (SCF)-method \citep{om68,hachisu86} to find the numerical solutions. The standard method is the following. Given the enthalpy $\hh$, the mass density is determined from Eq.(\ref{eq:eset2}b), then the gravitation potential from Eq.(\ref{eq:eset2}c). At this level, the three constants can then inferred from Eq.(\ref{eq:ctebc}). A new enthalpy is deduced from Eq.(\ref{eq:eset2}a) and one repeats these operations until input and output match (in the numerical sense). In other words, from a certain guess $\hh^{(0)}$, one gets a series of enthalpies $...,\hh^{(t-1)}, \hh^{(t)},...,\hh^{(\tf)}$, and the solution is found when $\hh^{(\tf)}$ is close enough to $\hh^{(\tf-1)}$. We see from Sect. \ref{eq:constants} that the triplets $(A,\hh_A)$, $(B,\hh_B)$ and $(M,\hh_M)$ are fixed points, i.e. these are preserved all along the SCF-cycle. Since point M where $\hh$ is maximum at convergence can not be guessed in advance, there are two options. Either this point is definitely held fixed, but the maximum of $\hh$ is never under control, or point M is allowed to change during the cycle \citep{om68,hachisu86}. With this second option, $\max (\hh)=1$ is easily guaranteed but a global re-scaling of the enthalpy field must be performed. This is straightforward with null external pressure because both $C_2$ and $C_3$ are independent on $\hh_M$. It is then sufficient to divide, at the end of each iteration, the enthalpy field $H$ by its largest value $\max (\hh)$. There are two by-products of this simple operation: first $\hh_M$ is assigned to constant $C_1$, and second, point M is the grid node where the actual maximum occurs. In contrast, when $\hp_e>0$ at points A and B, this {\it standard SCF-iteration scheme no longer works} because dividing $\hh$ by its maximal value also affects enthalpy values at these two reference points.

\subsection{An intra-loop re-scaling operator}

\begin{figure}
\includegraphics[width=8.4cm,bb=0 0 481 692,clip=true]{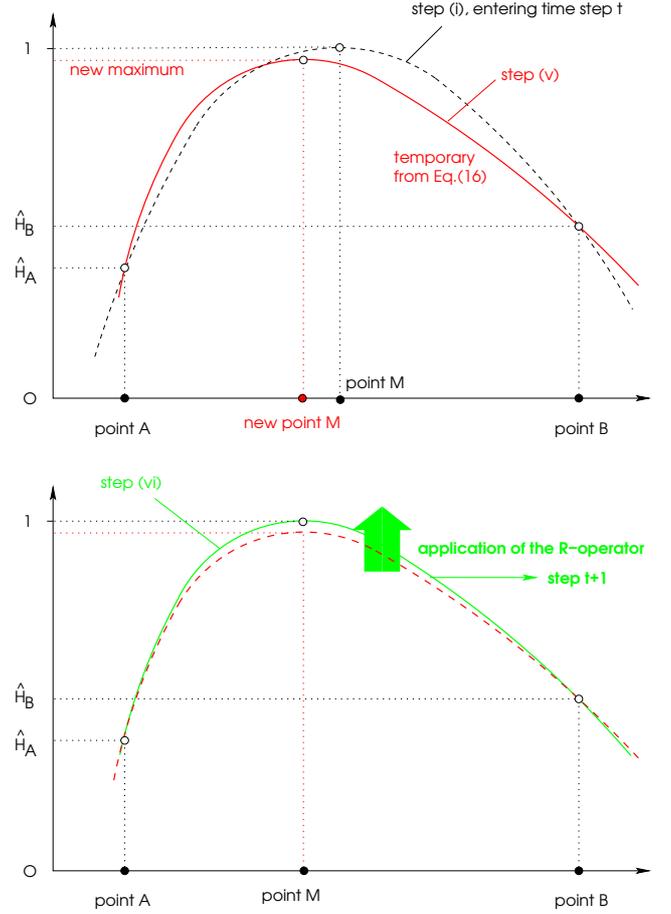}
\caption{Principle of the intra-loop re-scaling at a time step $t$ of the SCF-iterations. The example is given assuming that the three reference points are aligned. In general, enthalpy estimated at step (v) ({\it red}) is not unity at the actual point M. This is recovered by applying the re-scaling operator ${\cal R}$ at step (vi) ({\it green}). Not only values of $\hh$ at points A are B saved, but $\hmax$ remains under control.}
\label{fig:dynresc.eps}
\end{figure}

It follows from above that we need a specific mapping to converge towards self-normalised solutions, while respecting $\hh$ at the fixed points A and B. In the spirit of Lagrange interpolation, we define the re-scaling operator $\cal{R}$ by
\begin{flalign}
\label{eq:rescaledscf}
{\cal R}(\hh) = \ell_A(\hh) \hh_A +  \ell_{M}(\hh) \hnorm + \ell_B(\hh) \hh_B,
\end{flalign}
where the basis functions are
\begin{flalign}
\begin{cases}
\ell_A(\hh)=\frac{(\hh-\hh_M)(\hh-\hh_B)}{(\hh_A-\hh_M)(\hh_A-\hh_B)},\\
\ell_{M}(\hh)=\frac{(\hh-\hh_A)(\hh-\hh_B)}{(\hh_M-\hh_A)(\hh_M-\hh_B)},\\
\ell_B(\hh)=\frac{(\hh-\hh_M)(\hh-\hh_A)}{(\hh_B-\hh_M)(\hh_B-\hh_A)},
\end{cases}
\end{flalign}
where $\hh_A \equiv \hh_e$ at point A, and so on for $\hh_B$. It is easy to verify that this transformation leaves unchanged the enthalpy at points A and B, i.e. ${\cal R}(\hh_A)=\hh_A$ and ${\cal R}(\hh_B)=\hh_B$. Thus, if the starting guess $\hh^{(0)}$ has correct values at A and B, then these are preserved all along the cycle. Second, the formula enables to control the value of $\hh$ at a third floating point M, through the value $\hnorm$, which is a free parameter. Since $\ell_{M}(\hh_M)=1$, we actually have ${\cal R}(\hh_M)=\hnorm$ whatever $\hh_M$. To get a self-normalised enthalpy field, one has just to set $\hnorm=1$ once for all. This is illustrated in Fig. \ref{fig:dynresc.eps}. Another value will selected a different solution.

Interestingly enough, ${\cal R}(\hh)$ can be rewritten in the form
\begin{flalign}
\label{eq:rescaledscf_bis}
{\cal R}(\hh) = \hh + \frac{(\hh_M-\hnorm)(\hh-\hh_A)(\hh-\hh_B)}{(\hh_M-\hh_A)(\hh_M-\hh_B)},
\end{flalign}
which first indicates that the re-scaling is mainly dictated by the term $\hh_M - \hnorm$. This value can be large if both $\hmax$ and point M vary significantly during the iterations. Corrections are therefore expected to be the largest in the first steps of the iterative process until point M  where $\hh$ is maximum gets stable, and then softer as  $\hh_M - \hnorm \rightarrow 0$. Second, at convergence where $\hh$ no more evolves, the operator becomes exactly neutral everywhere, i.e. ${\cal R}(\hh) = \hh$. This property is essential. The accuracy of the solutions are therefore not impacted by applying the ${\cal R}$-operator.

\subsection{Special cases}
\label{subsec:sc}

When $\hh_B-\hh_A = 0$ but $\hh_A \ne 0$ (this includes the case of uniform external pressure), Eq.(\ref{eq:rescaledscf}) still works but must be rearranged. A linear re-scaling is, however, more straightforward in this case and the following choice
\begin{flalign}
{\cal R}(\hh) =\frac{(\hh-\hh_A)}{(\hh_M-\hh_A)}(\hnorm-\hh_A)+\hh_A
\end{flalign}
is sufficient. If $\hh_A = \hh_B = 0$ (which does not necessarily mean $\hh_e=0$ all along the boundary), we have
\begin{flalign}
{\cal R}(\hh) =\frac{\hh}{\hh_M}\hnorm,
\label{eq:rop}
\end{flalign}
which is the standard normalisation \citep{hachisu86}.

\section{Algorithm and numerical setup}
\label{sec:ans}

\subsection{The new algorithm}

We now summarise the main steps of the generalised SCF-method designed for over-pressured fluids. The equilibrium is obtained by computing successively, at step $t$
\begin{enumerate}
\item $\hrho^{(t)}$ from Eq.(\ref{eq:eset2}b),
\item $\hpsi^{(t)}$ from Eq.(\ref{eq:eset2}c),
\item $\hphi^{(t)}$ from the rotation law, 
\item constants $C_1^{(t)}$, $C_1^{(t)}$ and $C_3^{(t)}$ from Eqs.(\ref{eq:ctebc}),
\item a temporary enthalpy field $\hh^{\rm tmp}$ from Eq.(\ref{eq:eset2}a)
\begin{equation}
\hh^{\rm tmp} = \frac{C_3^{(t)}-C_2^{(t)} \hphi^{(t)}-\hpsi^{(t)}}{C_1^{(t)}}.
\label{eq:scfnewhtemph}
\end{equation}
\item the new, self-normalised enthalpy field
\begin{equation}
\hh^{(t+1)} = {\cal R}(\hh^{\rm tmp}).
\label{eq:scfnewh}
\end{equation}
\end{enumerate}
The difference with the standard SCF-algorithm stands at step (vi) where we have traditionally $\hh^{(t+1)} = \hh^{\rm tmp} /\hh^{(t)}_M $ with $\hh_M=\max (\hh)$, according to Eq.(\ref{eq:rop}).  The iterations begin from a guess $\hh^{(0)}$ and for a given pressure at the boundary $(\Gamma)$ and stop when $\hh$ gets stabilised (see below). Since this boundary is not known in advance, $\hh_e$ must be pre-defined in the entire computational space.

\subsection{Global numerical setup}
\label{subsec:code}

We have implemented the generalised SCF-method into the code {\tt DROP} described in Paper I. The results presented in the next sections are obtained for a computational box with $N=129$ grid nodes per direction, corresponding to $7$ levels of multigrid. In contrast with Paper I devoted to rings, the radial and vertical spacings are allowed to differ from each other, in order to optimise the covering factor $\Lambda$ (fluid section-to-grid area ratio). This is efficient for oblate shapes mainly. The expected accuracy of most quantities is of the order of $10^{-4}$ typically, which is indeed observed, in particular for the Virial parameter (see below).

\subsection{Seed and convergence criterion}

The SCF-method is capable of converging to a given solution for a broad variety of seeds $\hh^{(0)}$. However, if $\hh^{(0)}$ is too far from the target, then algorithm can fail or converge to something very different. Typically, $\hpsi$ has a paraboloidal shape, at least in the vicinity of the mass distribution, and it is symmetric with respect to the equatorial plane. A good guess is an enthalpy of the form
\begin{equation}
\hh^{(0)}=a(\hr-\hr_c)^2+b\hz^2+1
\end{equation}
where coefficient $\hr_c$, $a$ and $b$ and are easily deduced once the two reference points A and B and associated enthalpies are defined (see Fig. \ref{fig:tore_et_ell.eps}). We are lead to the following values
\begin{flalign}
  \begin{cases}
    \hr_c=0,\\
    a=\frac{\hh_B-1}{\hr_B^2},\\
b=\frac{\hh_A-1}{\hz_A^2},\\
\end{cases}
\label{eq:hguesse}
\end{flalign}
for ellipsoidal configurations, where point A stands on the rotation axis, and point B is the equatorial radius. For rings, we set
\begin{flalign}
  \begin{cases}
    e=\sqrt{\frac{1-\hh_A}{1-\hh_B}},\\
\hr_c= \frac{\hr_A+e \hr_B}{1+e},\\
a=\frac{\hh_B-1}{(\hr_B-\hr_c)^2},\\
b=2\frac{\hh_B+\hh_A-2}{(\hr_B-\hr_A)^2},\\
\end{cases}
\label{eq:hguesst}
\end{flalign}
where points A and B are the inner and outer edges, respectively.

Regarding the convergence criterion, there are different possibilities. As in Paper I, we work with the Euclidean norm $||\delta \hh^{(t)}||_F =\hh^{(t)} -\hh^{(t-1)}$. Since the computational box has $N+1$ nodes per directions, the SCF-iterations are stopped as soon as
  \begin{equation}
    \frac{1}{(N-1)^2}|| \delta \hh^{(t)}||_F = \lesssim \epsilon
  \end{equation}
where $\epsilon$ is of the order of the computer precision, and $|| \delta \hh^{(t)}||_F$ accounts for interior points only.

\section{Results for rigid rotation and uniform external pressure}
\label{sec:uniformp}

In this section, we set $\ho(\hr)=1$ corresponding to rigid rotation and $\hh_e(\vec{r}_\Gamma)= \in [0,1[$ (see below for differential rotation).

\subsection{Output quantities}
\label{subsec:oq}

From the converged enthalpy $\hh$, constants $C_1$, $C_2$ and $C_3$, all global quantities can be deduced, namely (in dimensionless form): the area of the section $\hat{S}$, the fluid volume $\hat{V}$, the mass $\hat{M}$, the mean mass density $\langle \hrho \rangle = \hm/\hat{V}$, the angular momentum $\hjcin$, the gravitational energy $\hw$, the integral of pressure $\hpi$, the internal energy $\hu=3\hpi$ and kinetic energy $\hcin$ (see paper I). The quality of the numerical solution depends on many factors (grid resolution and order of schemes mainly). It can be checked by considering the Virial equation
 \citep{cox1968}
\begin{equation}
W+2T+U-\oint{P_e \vec{r} \cdot d\vec{A}}=0,
\label{eq:virielwextp}
\end{equation}
where $d\vec{A}=dA \vec{e}_\Gamma$ is the elementary area at the fluid surface, $\vec{e}_\Gamma$ is a unit vector oriented outward (see Fig. \ref{fig:tore_et_ell.eps}). The last term is the contribution of external pressure. The dimensionless version of this expression is
\begin{equation}
\hw+\frac{C_1}{n+1}\left(3\hpi-\hpi_e\right)+2 C_2 \hcin =0,
\label{eq:virielwextp_dedim}
\end{equation}
where
\begin{flalign}
\hpi_e= \oint{\hp_e \cos(\vec{e}_r,\vec{e}_\Gamma) \hrs d\hba}.
\end{flalign}
This integral over the fluid surface is just $3\hp_e \hvol$ if the external pressure is uniform. Because of axial symmetry, this is finally a one-dimension integral. We have $d\hba=2 \pi \hr d\hat{s}$ and then
\begin{flalign}
\hpi_e= 2 \pi \oint_\Gamma{\hp_e \cos(\vec{e}_r,\vec{e}_\Gamma) \hrs \hr d\hat{s}},
\label{eq:pie}
\end{flalign}
where $\hat{s}$ is a curvilinear coordinate along ($\Gamma$). By dividing  Eq.(\ref{eq:virielwextp_dedim}) by the largest term $|\hw|$, the relative Virial parameter is
\begin{equation}
VP \equiv -\frac{1}{\hw}\left[\frac{C_1}{n+1}\left(3\hpi-\hpi_e\right)+2 C_2 \hcin \right] -1.
\label{eq:rvp}
\end{equation}
 The lower this number, the more self-consistent the solution. The main sources of error are the numerical solution $\hh$ and the quadrature leading to $VP$. As argued in Paper I, $VP$ is typically of the order of $1/(N+1)^2$ for second-order schemes.

\begin{figure}
  \centering
  \includegraphics[height=5.1cm,bb=76 270 483 675, clip==]{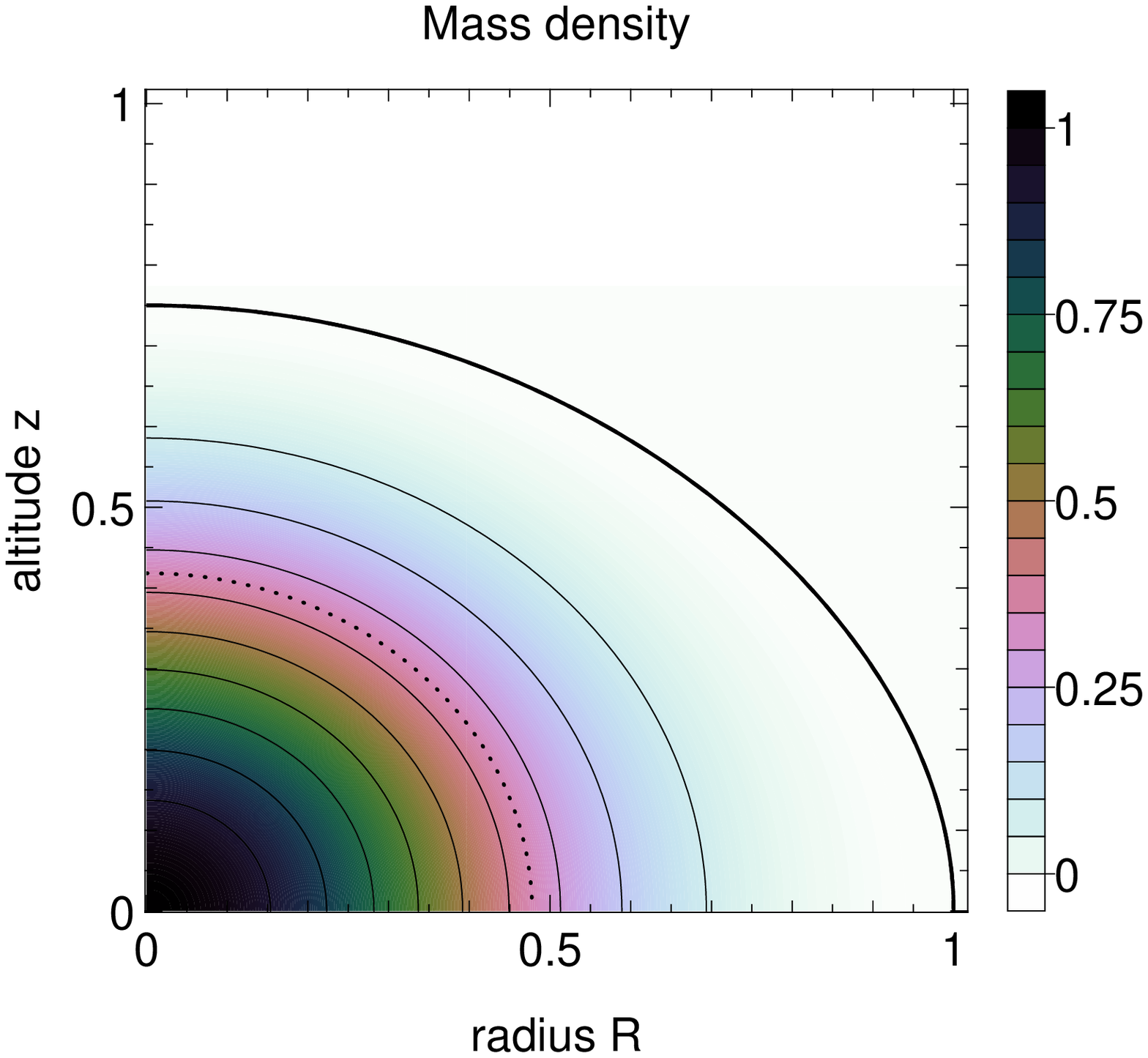}\\
  \bigskip
  \hspace*{25pt}\includegraphics[height=5.1cm,bb=76 270 553 675,clip=true]{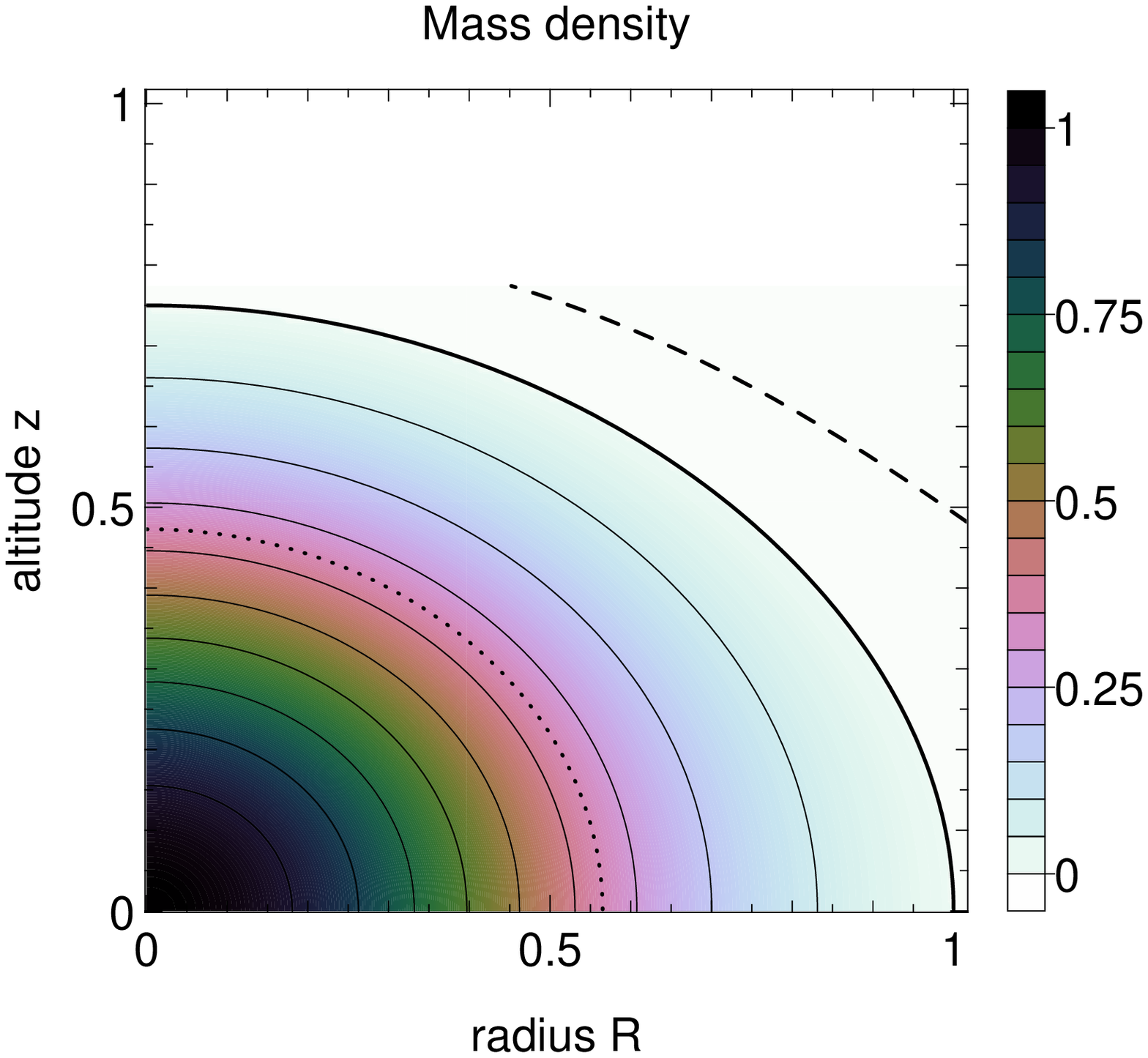}
\caption{Mass density structure for a uniformly rotating ellipsoid with polytropic index $n=1.5$ and axis ratio $\hz_A/\hr_B=0.75$ in the isolated case ({\it left}) and with uniform external pressure corresponding to $\hh_e=0.1$ ({\it right}). Density contours are every $\Delta \hrho=0.1$ ({\it thin lines}). Also shown are the fluid boundary where $\hh - \hh_e=0$ ({\it bold}), the zero enthalpy level ({\it dashed}), and the place where $\hh=0.5$ ({\it dotted}), which corresponds to $\hrho \approx 0.353$. Here, the top of the computational box stands at $z \sim +0.774$.}
\label{fig:config_overp_ell.eps}
\end{figure}

\begin{table}
\centering
\begin{tabular}{llll}\\
  quantity                  & \cite{hachisu86}    & $\hh_e=0$   & $\hh_e=0.1$ \\ \hline
  covering factor $\Lambda$ & $0.091^*$            & $0.732$   & $  0.738$  \\
  $C_1$                     & $0.625^*$             & $0.6250$ & $  0.8293$   \\
  $C_2$ (i.e. $\Omega_0^2)$  & $0.227$              & $0.2268$ & $  0.2954$   \\
  $-C_3$                     & ?                     & $0.5504$ & $  0.7548$    \\
  $\hr_e$                    &  $1$                & $1$ & $1$\\
  $\hr_p$                    &  $  0.75$           & $  0.75$ &$  0.75$\\
 $\hbsur$                   &  ?                   & $0.1515$ & $  1.1606$\\
  $\hv$                      & $3.03$             & $3.0291$ & $  3.0684$ \\
  $\hm$                      & $0.430$             & $0.431$ & $  0.6726$\\
  $\langle \hrho \rangle$    & $0.142^*$           & $0.1422$ & $ 0.2192$\\
  max. pressure              & $0.250$             & $0.2500$ &$  0.3317$\\
  max. density               & $0.494^*$          & $0.4941$ & $  0.7552$\\
  $\sqrt{C_2}\hjcin$         & $0.0562$             & $0.0356$ & $  0.0868$\\
  $C_2 \hcin$                & $0.00847$             & $0.00848$ & $  0.0236$\\
  $-\hw$                     & $0.183$             & $0.18378$ & $  0.3882$\\
  $\beta$                    & $0.046$              & $0.051$  &  $0.061$ \\
  $\frac{C_1}{n+1}\hu$       & $0.167$             & $0.16682$ & $  0.3507$\\
  $-\Pi_e$                   & $0$                  &$0$        & $ 0.0097$\\
  $\log(VP)$                 & ?                  & $-5.04$ &   $ -3.98$\\
  iterations & ?        & $38$   &  $ 29$\\\hline
  $^*$estimated.\\
\end{tabular}
\caption{Results for the equilibria shown in Fig. \ref{fig:config_overp_ell.eps} (see Sect. \ref{subsec:code} for the numerical setup). The last column is for the over-pressurized fluid. }
\label{tab:datae}
\end{table}

\subsection{Ellipsoids}
\label{subsec:ellipsoid}

The first example is an ellipsoid with index $n=1.5$ and axis ratio $0.75 \equiv \hz_A/\hr_B$. Figure \ref{fig:config_overp_ell.eps} displays the mass density at equilibrium computed for $\hh_e(\Gamma)=0.1$, which corresponds to a relatively weak pressure contrast $\sim 0.0032$ between the boundary and the core of the fluid. The structure obtained in the isolated case is also shown in comparison. Due to a significant rotation rate, the line where $\hh=0$ is located well outside the fluid. It is not homothetical with $(\Gamma)$. We clearly see that the core slightly expands because $\hrho > 0$ on the fluid boundary. We give in Tab. \ref{tab:datae} output quantities for this run. As anticipated before, external pressure increases the dimensionless mass and the rotation rate. The volume is very weakly increased. The fluid is in fact a little bit thicker (in the vertical direction), due to the deeper potential well. We notice that the Virial parameter is globally very good with about $4$ correct digits at the actual resolution.

\begin{figure}
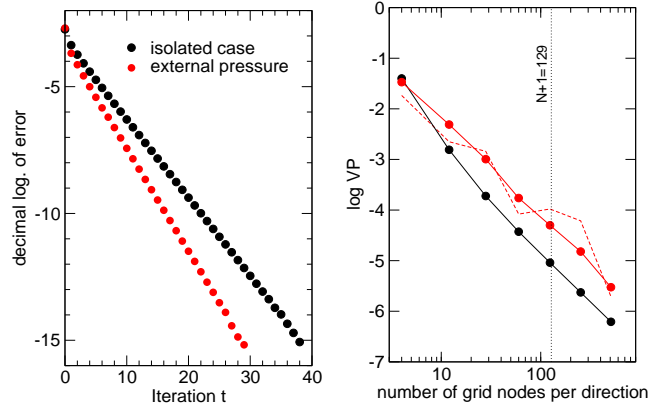

      \centering
  \includegraphics[width=4.1cm,bb=27 46 400 522,clip=true]{conv_ell.eps}\quad\includegraphics[width=4.cm,bb=39 41 404 530,clip=true]{vp_ell.eps}
\caption{Evolution of convergence during the SCF-iterations for the ellipsoid considered in Fig. \ref{fig:config_overp_ell.eps} ({\it left}), and Virial parameter versus the number $N+1$ of grid nodes per direction ({\it right}). The isolated case is shown in comparison ({\it black}). The relationship between $\log VP $ and $\log(N+1)$ is stricty linear when the sub-grid approach is implemented ({\it dotted lines}).}
\label{fig:conv_ell.eps}
\end{figure}

The convergence of the SCF-iterations is displayed in Fig. \ref{fig:conv_ell.eps}a. Much less iterations than in the isolated case are required. This is not really a surprise since external pressure flattens the mass density profile (and $\tf$ rises with $n$; see Paper I). The variation of $\log(VP)$ with $\log(N+1)$ is displayed in Fig. \ref{fig:conv_ell.eps}b. Without subgrid approach (see paper I), the linear behaviour is not strictly observed and some wiggles appear.

We have repeated the simulation for a series of enthalpy values $\hh_e$ in the range $[0,1[$. The fluid response to over-pressurization is linear only $\hh_e \lesssim 0.1$ typically. This depends strongly on $n$ (see below). Constants $C_1$ and $C_2$ vary as expected. In particular, $C_2$ asymptotically attains the value corresponding to the Maclaurin fluid. The evolution of constant $C_3$ (not shown) is not trivial, which is consistent with the discussion in Sec. \ref{subsec.wcwe}. It is first negative and decreases, then goes through a minimum at $\hh_e \approx 0.375$, i.e. $\hp_e \approx 0.087$, then increases. It vanishes for  $\hh_e \approx 0.68$ ($\hp_e \approx 0.38$) and diverges as $\hh_e \rightarrow 1$. The volume and section area are, however, essentially not impacted.

\begin{figure}
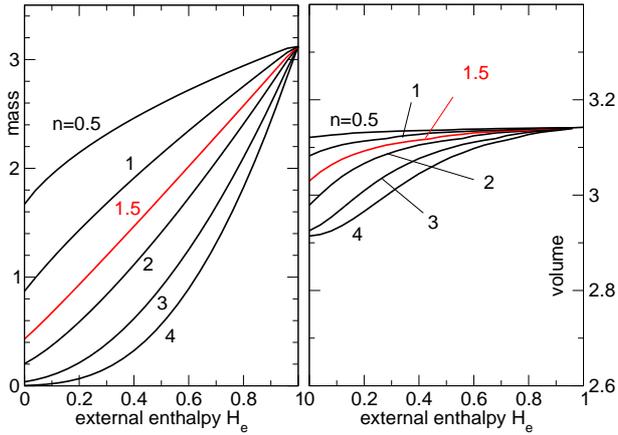

  \centering
  \includegraphics[width=3.9cm,bb=72 38 403 522,clip==]{mass_he_ell.eps}\includegraphics[width=4.12cm,bb=85 38 434 522,clip==]{volume_he_ell.eps}\caption{Dimensionless mass $\hm$ ({\it left}) and volume $\hv$ ({\it right}) as a function of the external enthalpy $\hh_e$ for an ellipsoid with axis ratio $0.75$. The polytropic index $n$ is labelled on the curves.}
\label{fig:massvol_he_ell.eps}
 \end{figure}

\begin{figure}
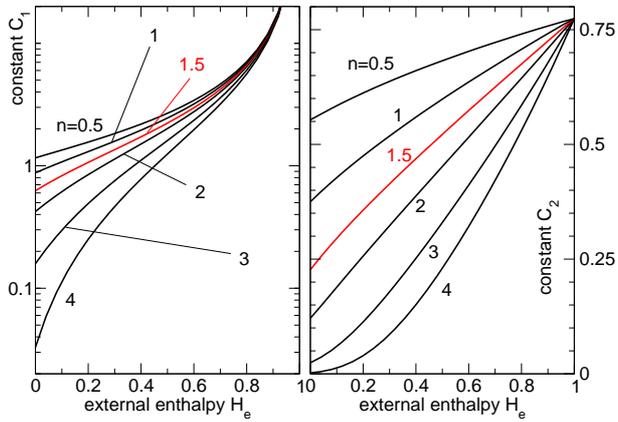

  \centering
  \includegraphics[width=3.91cm,bb=59 38 403 523,clip=]{C1_he_ell.eps}\includegraphics[width=4.1cm,bb=85 38 446 522,clip=]{C2_he_ell.eps}
\caption{Same legend as for Fig. \ref{fig:massvol_he_ell.eps} but for constants $C_1$ ({\it left}) and $C_2$ ({\it right}).}
\label{fig:c1c2_he_ell.eps}
\end{figure}

\subsection{Varying the polytropic index}
 
We have considered various polytropic indices from $n=0.5$ up to $4$. The axis ratio and numerical setup are unchanged. The results are displayed in Fig. \ref{fig:massvol_he_ell.eps} for the dimensionless mass and volume and in Fig. \ref{fig:c1c2_he_ell.eps} for the two constants $C_1$ and $C_2$. We see that the larger the index $n$, the smaller $C_1$, $C_2$, $\hm$ and $\hv$, while the amplitudes of variations are larger for larger indices. In contrast with $\hh_e$, $\hrho$ is very sensitive to $n$: the larger $n$, the more peaked the mass density and the larger the wings. Because the reference points are held fixed, the geometrical quantities (section area and volume) do not vary significantly. All curves converge to the Maclaurin values as $\hh_e \rightarrow 1$, as expected (see Sec. \ref{subsec.pbatfb}).

\begin{figure}
  \centering
  \includegraphics[height=5.1cm,bb=76 270 483 675, clip==]{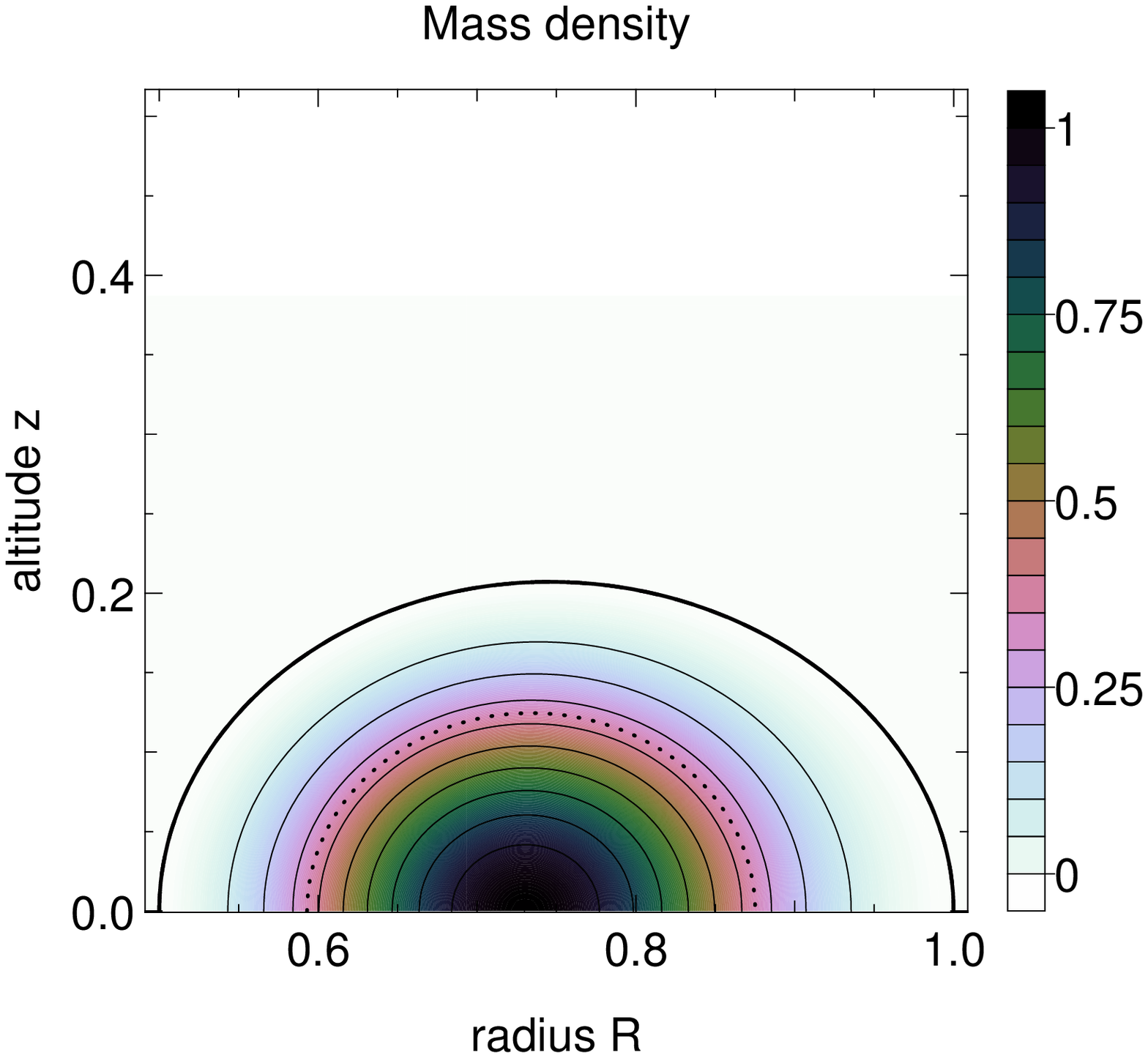}\\
  \bigskip
  \hspace*{25pt}\includegraphics[height=5.1cm,bb=76 270 553 675,clip==]{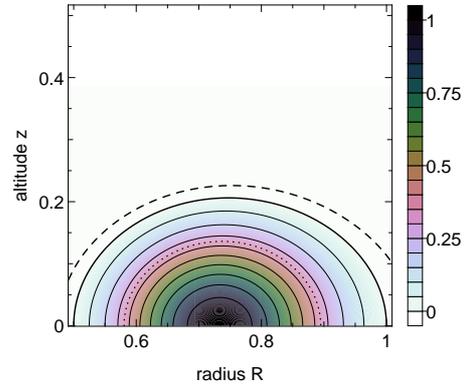}
\caption{Same legend/conditions as for Fig. \ref{fig:config_overp_ell.eps} but for a ring with axis ratio $\hr_A/\hr_B=0.5$.}
\label{fig:config_overp_torus.eps}
\end{figure}

\subsection{Rings}
\label{subsec:torus}

As a second example, we consider a torus/ring with index $n=1.5$ and axis ratio $\hr_A/\hr_B=0.5$. Again, external pressure is maintained uniform in the free space, and subsequently on the converged fluid boundary. The mass density stratification  obtained for $\hh_e=0.1$ is displayed in Fig. \ref{fig:config_overp_torus.eps}. The solution converges faster than for zero exterior stress by a factor $30 \%$ typically. Output quantities are listed in Tab. \ref{tab:datat}. The Virial parameter remains very good. We clearly see, as for the ellipsoid, the slight extension of the core. Besides, $\hm$ and $C_2$ are enhanced, which increases the componenents of the Virial equation.

  We have analysed the effect of varying the external pressure (i.e. $\hh_e$), while holding $n$ and $\hr_A/\hr_B$ fixed. Similarities with the ellipsoids studied above are striking. Again, the section area $\hbsur$, polar radius $\hr_p$ and volume $\hv$ are almost conserved. The dimensionless mass increases (due to a flatter mass density profile) by a factor about $3$ between the isolated case and the maximal pressure case. Constant $C_1$ gradually diverges as $\hh_e$ increases, while $C_2$ reaches $0.54$. Again, $C_3$ goes through a minimum, now located at $\hh_e \approx 0.6$, i.e. $\hp_e \approx 0.28$, and vanishes for $\hh_e \approx 0.9$ ($\hp_e \approx 0.77$).

\subsection{Varying the polytropic index}

\begin{figure}
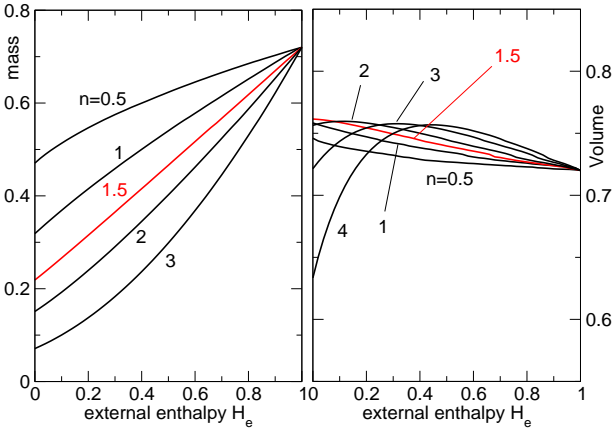

  \centering
\includegraphics[width=4.cm,bb=55 38 403 530,clip=]{mass_he_torus.eps}\includegraphics[width=4.02cm,bb=85 38 434 522,clip=]{volume_he_torus.eps}
\caption{Dimensionless mass ({\it left}) and volume ({\it right}) for a torus with axis ratio $0.5$ as a function of the external enthalpy $\hh_e$, and for different polytropic indices.}
\label{fig:massvol_he_torus.eps}
\end{figure}

\begin{figure}
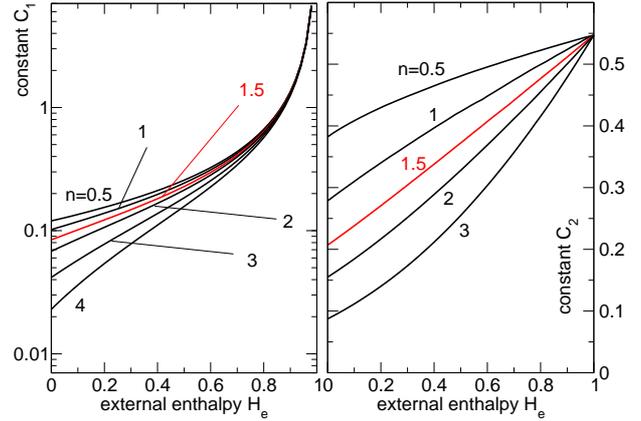

  \centering
  \includegraphics[width=4.07cm,bb=47 38 403 522,clip=]{c1_he_torus.eps}\includegraphics[width=4.cm,bb=85 38 434 522,clip=]{C2_he_torus.eps}
\caption{Same legend as for Fig. \ref{fig:massvol_he_torus.eps} but for the two constants $C_1$ ({\it left}) and $C_2$ ({\it right}).}
\label{fig:c1c2_he_torus.eps}
\end{figure}

Figure \ref{fig:massvol_he_torus.eps} shows the variation of dimensionless mass and volume with $\hh_e$ for a few polytropic indices $n$. The two constants $C_1$ and $C_2$ are given in Fig. \ref{fig:c1c2_he_torus.eps}. Globally, we find for rings the same trends as for ellipsoids. As $\hh_e \rightarrow 1$, $C_1$ diverges, the fluid section tends to an infinitely thin, circular loop. There is a slight non-monotonic effect on the volume: for indices $n \lesssim 2.5$ typically $d\hv / d\hh_e >0$, while for larger indices, the volume first increases with increasing $\hp_e$, and then decreases.

\begin{table}
\centering
\begin{tabular}{llll}\\
  quantity                  & \cite{hachisu86}    & $\hh_e=0$   & $\hh_e=0.1$ \\ \hline
  covering factor $\Lambda$ & $0.091^*$            & $0.608$   & $0.605$   \\
  $C_1$                     & $0.0842^*$             & $0.08422$ & $  0.1016$   \\
  $C_2$ (i.e. $\Omega_0^2)$  & $0.207$              & $0.2068$ & $  0.2382$    \\
  $-C_3$                     & ?                     & $0.3691$ & $  0.4339$     \\
  $\hr_e$                    &  $0.5$              & $0.5$ & $0.5$\\
  $\hr_p$                    &  ?                   & $0.2071$ &$  0.2061$\\
 $\hbsur$                   &   ?                  & $0.1619$ & $  0.1612$\\
  $\hv$                      & $0.762$             & $0.7615$ & $  0.7588$ \\
  $\hm$                      & $0.219$             & $0.2188$ & $  0.2661$\\
  $\langle \hrho \rangle$    &  $0.287^*$           & $0.2874$ & $ 0.3507$\\
  max. pressure              & $0.0337$             & $0.03369$ &$ 0.0406$\\
  max. density               &  $0.0245^*$          & $0.02444$ & $ 0.0324$\\
  $\sqrt{C_2}\hjcin$         & $0.0562$             & $0.05613$ & $ 0.0746$\\
  $C_2 \hcin$                & $0.0128$             & $0.01276$ & $ 0.0182$\\
  $-\hw$                     & $0.0401$             & $0.04005$ & $ 0.0575$\\
  $\beta$                    & $0.32$              & $0.32$  &  $0.32$ \\
  $\frac{C_1}{n+1}\hu$       & $0.0145$             & $0.01453$ &  $ 0.0214$\\
  $-\Pi_e$                   & $0$                  &$0$        & $0.000292$\\
  $\log(VP)$                 & ?                  & $-4.42$ &   $ -4.10$\\
  iterations & ?        & $65$   &  $ 44$\\\hline
  $^*$estimated.\\
\end{tabular}
\caption{Results for the equilibria shown in Fig. \ref{fig:config_overp_torus.eps}. The last column is for the over-pressurized fluid.}
\label{tab:datat}
\end{table}

\begin{figure*}
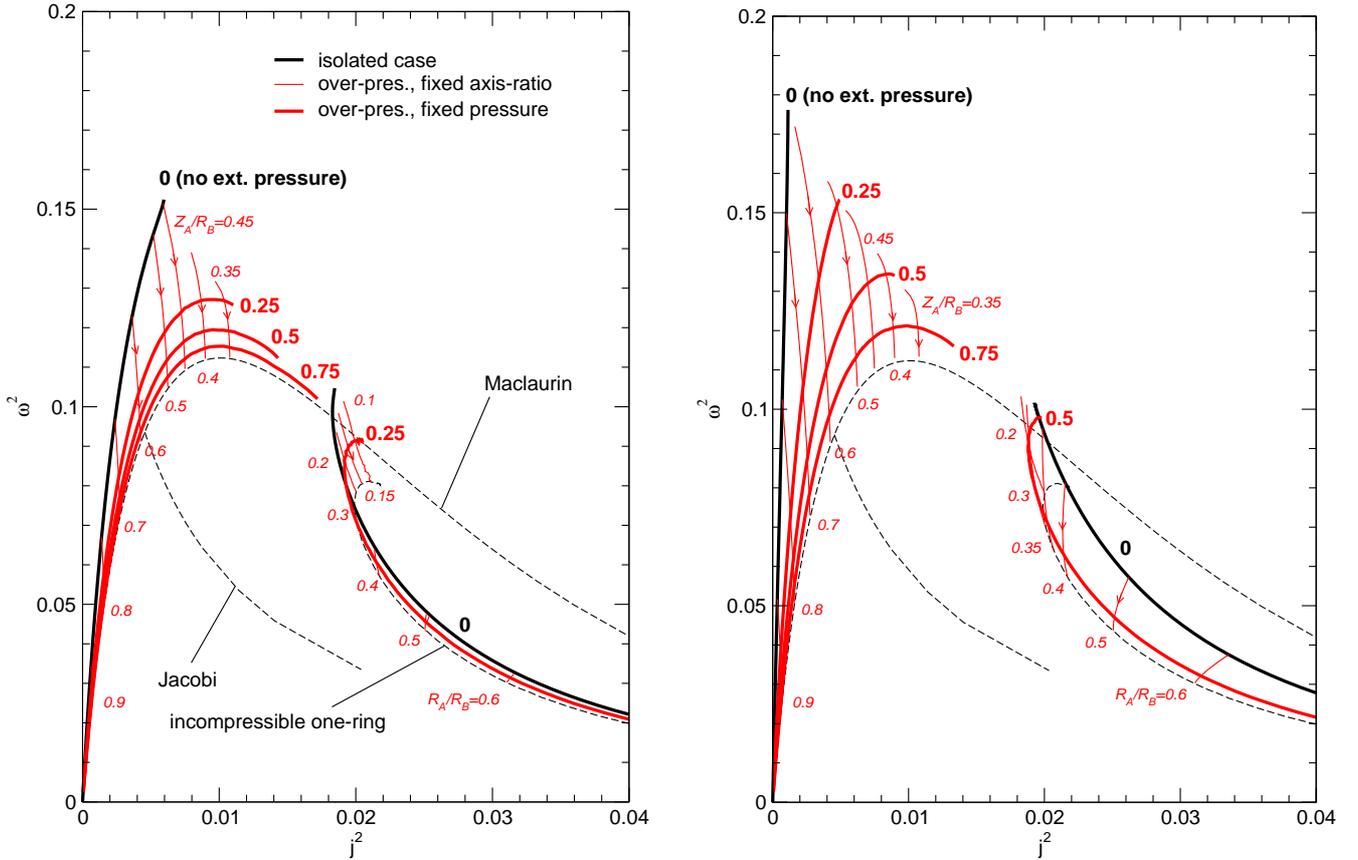

  \centering
  \includegraphics[width=8.5cm,bb=25 43 569 772,clip=]{hachisu_he_n05.eps}\qquad\includegraphics[width=8.5cm,bb=26 43 573 772,clip=]{hachisu_he_n15.eps}
\caption{Equilibrium sequences obtained for a self-gravitating polytrope over-pressurized by external photons: constant axis ratio with increasing $\hh_e$ ({\it red thin}), and fixed  $\hh_e$ with decreasing axis ratio ({\it red bold}).  Rotation is uniform (i.e. solid). Uniform pressure is assumed at the fluid boundary. The polytropic index is $n=0.5$ ({\it left}) and $n=1.5$ ({\it right}). The incompressible Maclaurin, one-ring and Jacobi sequences are shown in comparison ({\it dashed lines}).}
\label{fig:hachisu_he_n05.eps}
\end{figure*}

\begin{figure*}
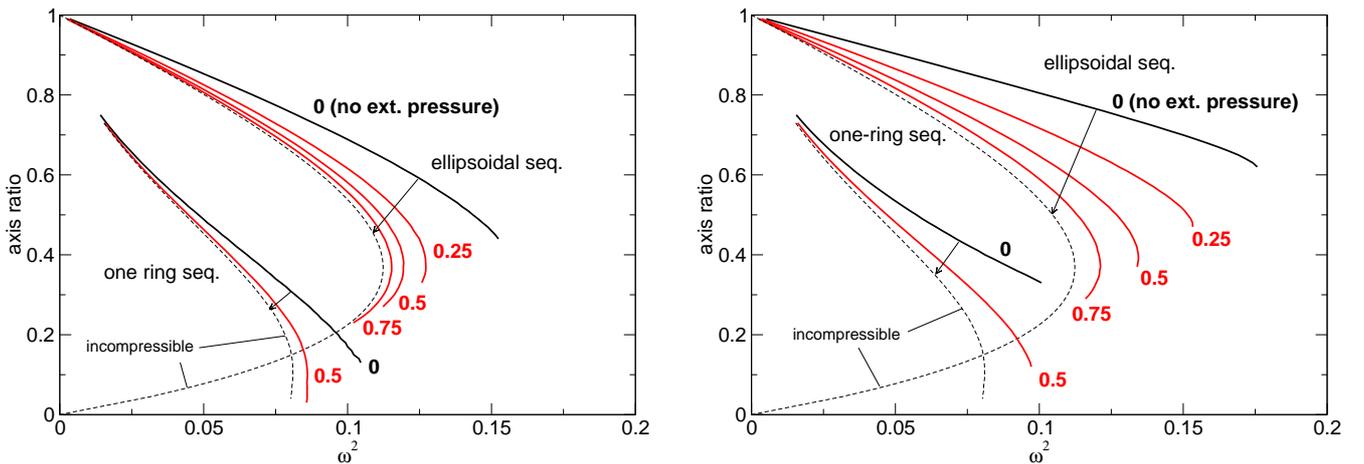

\includegraphics[width=8.5cm,bb=33 39 720 529,clip=]{evsomega2_0.5.eps}\qquad\includegraphics[width=8.5cm,bb=33 39 720 529,clip=]{evsomega2_1.5.eps}
\caption{Axis ratio ($\hz_A/\hr_B$ for ellipsoids and $\hr_A/\hr_B$ for rings) as a function of $\omega^2$ for over-pressurized polytropes ({\it red}). The polytropic index is $n=0.5$ ({\it left}) and $n=1.5$ ({\it right}). Values of $\hh_e$ are indicated on the curves, including the zero-pressure case ({\it black}). The incompressible branches are shown in comparison ({\it dashed lines}). See also Fig. \ref{fig:hachisu_he_n05.eps}.}
\label{fig:axis-ratio_he_n05.eps}
\end{figure*}

\section{Equilibrium sequences}
\label{eq:eseq}

\subsection{Location in the $\omega^2-j^2$ diagram}

A classical manner to visualise and compare equilibria obtained for different triplets $(C_1,C_2,C_3)$ is the $\omega^2-j^2$ diagram \citep{chandra73,hachisu86}, where
\begin{equation}
\begin{cases}
j^2 =\frac{1}{4\pi G \langle \rho \rangle}\frac{J^2}{M^2 V^{4/3}},\\
\omega^2 = \frac{1}{4\pi G \langle \rho \rangle}\Omega_0^2,
\end{cases}
\end{equation}
and $J = \rho_0 \Omega_0 L^5 \hjcin$. Using dimensionless quantities, we have
\begin{equation}
(j^2,\omega^2) = \frac{C_2}{4\pi \hm} \times \left(\frac{\hjcin^2}{\hm^2 \hv^{1/3}},\hv\right).
\label{eq:omega2j2}
\end{equation}
A diagram is obtained by varying the fluid aspect ratio, while holding the polytropic index fixed. For homogeneous ellipsoids, it is degenerate, with two possible states (spheroidal or very flat) for a given rotation rate $\omega$. For the one-ring sequence, $\omega$ mainly decreases as the ring becomes thinner and thinner, while $j$ increases \citep{hachisu86, ansorg03}. One goes continuously from one sequence to the other through pinched confugurations. As soon as $n>0$, equilibria are not always possible beyond some critical axis ratios. This depends on the index $n$ and rotation profile \cite[e.g.][]{hachisu86}. The same is expected here. We have constructed the $\omega^2-j^2$ diagram in the presence of external pressure for both ellipsoidal and toroidal configurations. We still work under the assumption of constant external stress. We have proceeded in two ways: i) the axis ratio varies (as traditionally done) while $\hh_e$ is fixed, and ii) $\hh_e$ varies and the axis ratio is fixed. Figure \ref{fig:hachisu_he_n05.eps} displays the graphs so obtained for $n=0.5$ and $n=1.5$. The Maclaurin sequence has been added, as well as the one ring and Jacobi sequences that branches off at $\hz_A/\hr_B =0$ and $\sim 0.58$ respectively. We see that the new ellipsoidal sequences are located in between the Maclaurin sequence and the zero-external pressure branch. There is a perfect connection between these two sequences when external pressure varies continuously in the range $[0,1]$ for all axis ratios larger than the critical axis ratios, approximately $0.44$ for $n=0.5$ and $0.32$ for $n=1.5$. The sequence endings move largely. This is somewhat similar for the ring sequences. Near critical rotations, however, the new ring solutions cross the incompressible branch. This is clearly a zone of non-linearity. Globally, we observe that the ellipsoidal and toroidal sequences get closer and closer to the incompressible branches as $\hh_e$ increases. This is fully expected (see Sect. \ref{subsec.wcwe}).

\subsection{Critical rotations exceeded}

\begin{figure}
   \centering
  \includegraphics[height=5.1cm,bb=76 270 483 675, clip==]{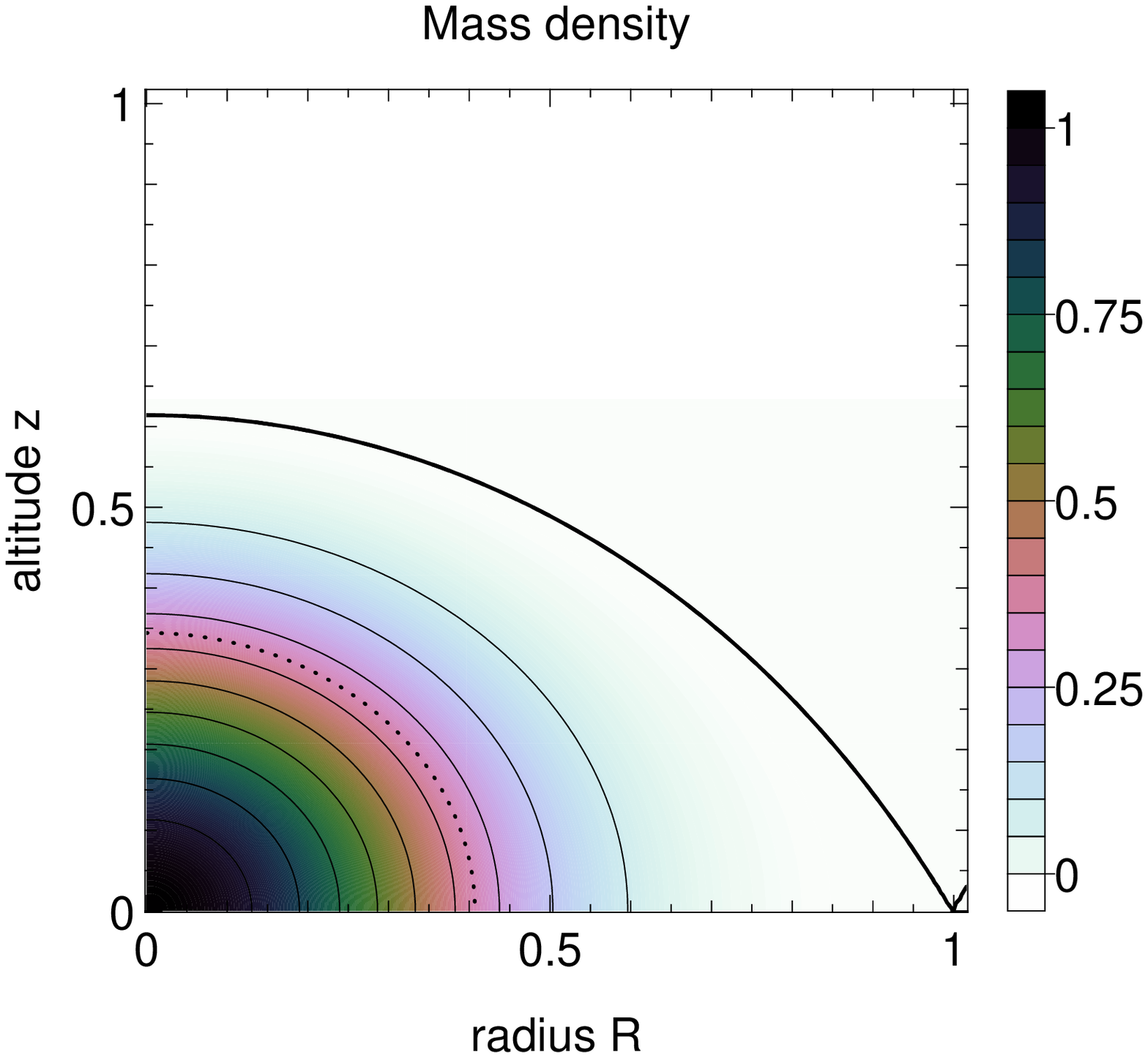}\\
  \bigskip
  \hspace*{25pt}\includegraphics[height=5.1cm,bb=76 270 553 675,clip=true]{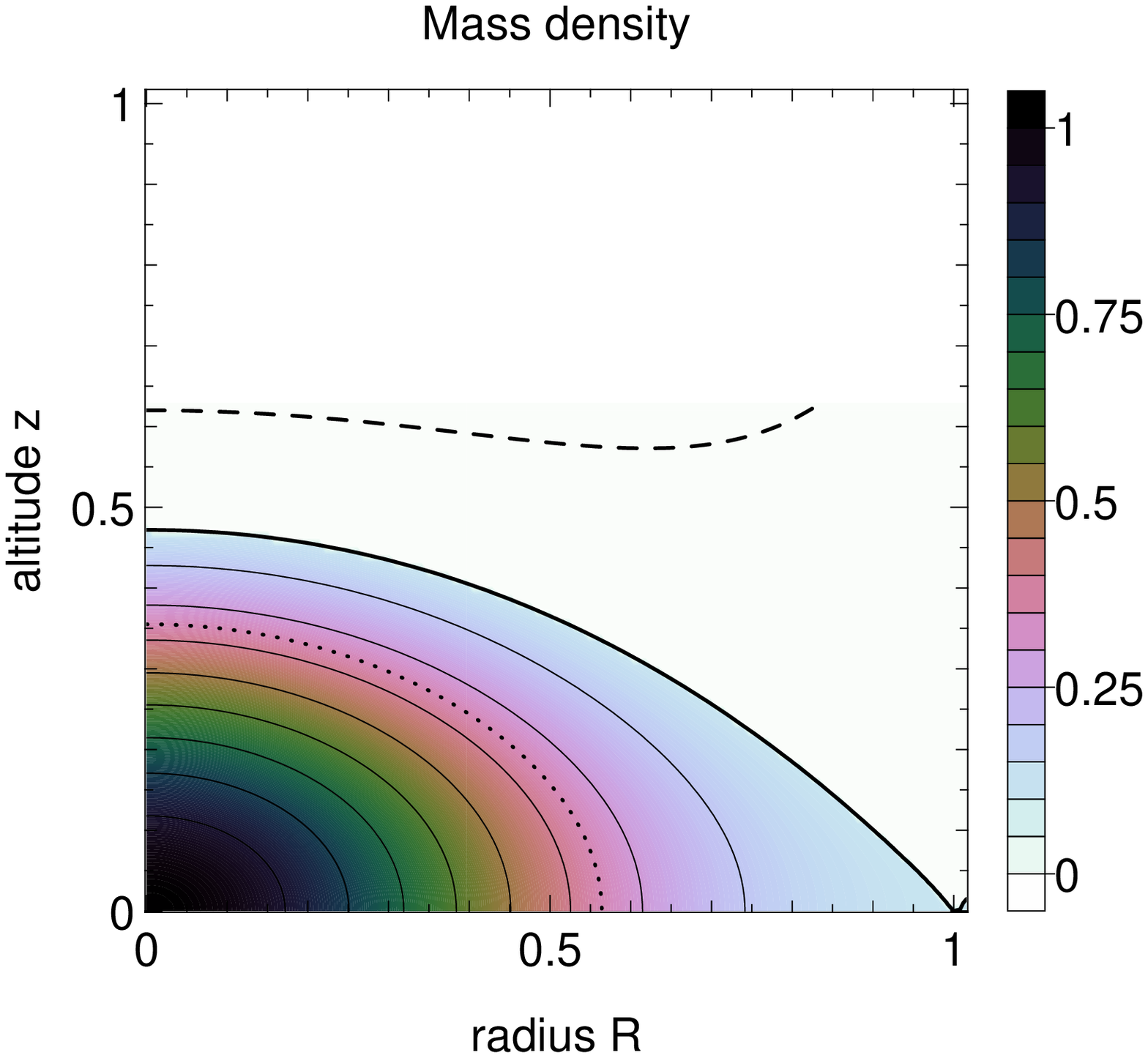}
\caption{Mass density structure for the states of critical rotation of the ellipsoidal sequence with $n=1.5$, without external pressure ({\it top}) and for $\hh_e=0.25$ ({\it bottom}).  Density contours are every $\Delta \hrho=0.1$ ({\it thin lines}). Also shown are the fluid boundary where $\hh - \hh_e=0$ ({\it bold}), the zero enthalpy level ({\it dashed}), and the place where $\hh=0.5$ ({\it dotted}), which corresponds to $\hrho \approx 0.353$. See also Tab. \ref{tab:cr}.}
\label{fig:crot_overp_ell.eps}
\end{figure}

\begin{figure}
  \centering
  \includegraphics[height=5.1cm,bb=76 270 483 675, clip==]{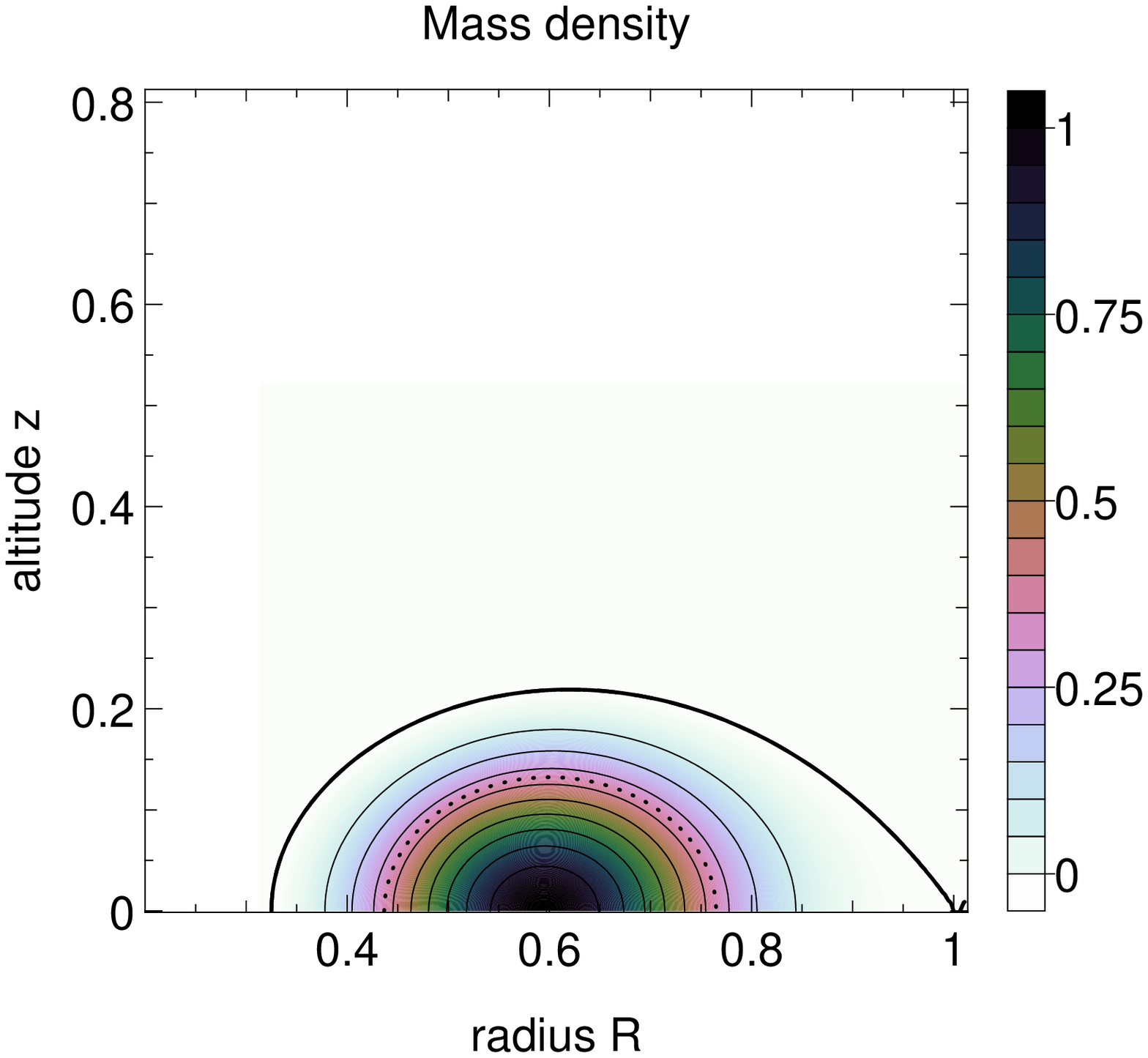}\\
  \bigskip
  \hspace*{25pt}\includegraphics[height=5.1cm,bb=76 270 553 675,clip=true]{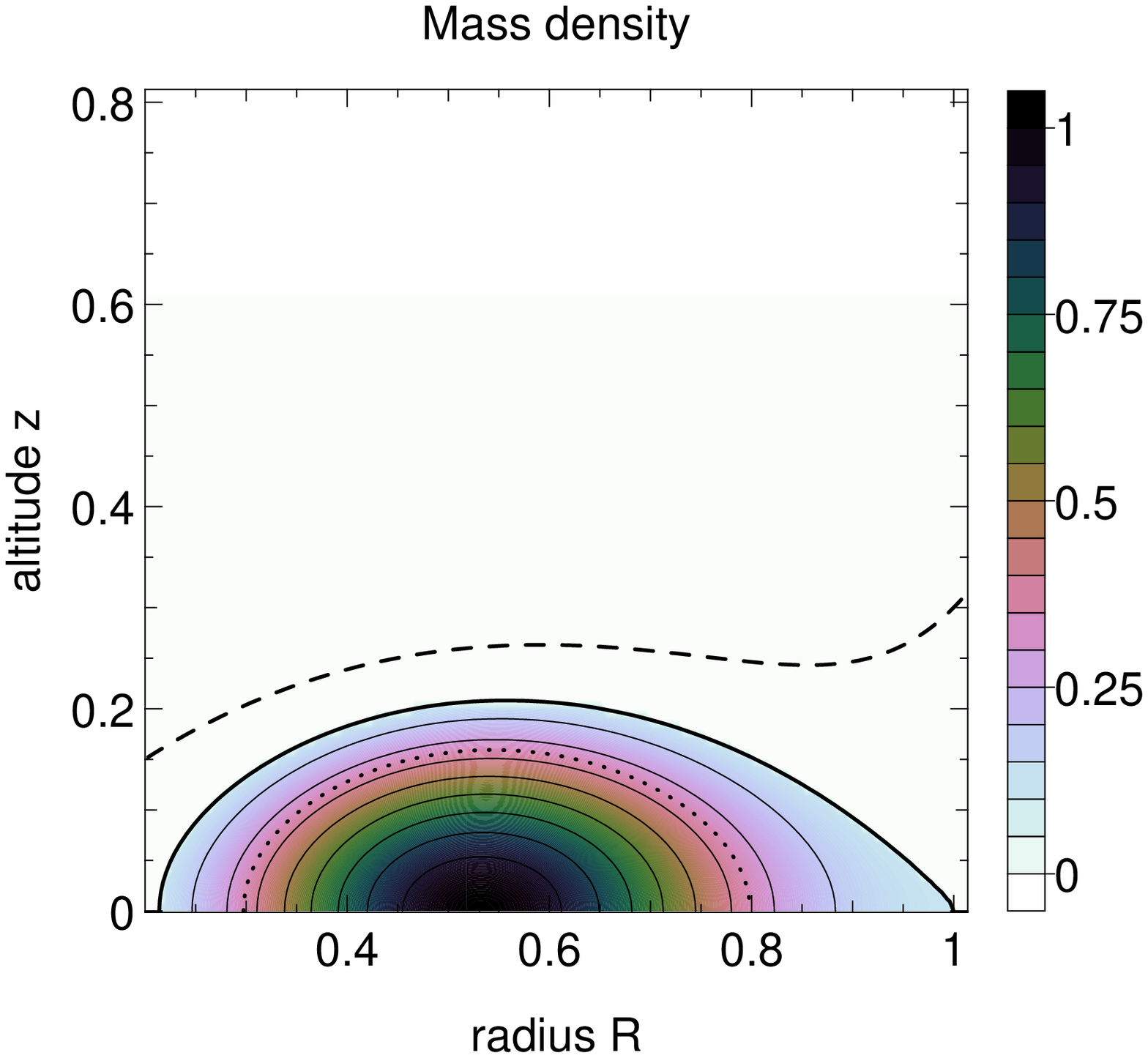}
\caption{Same legend as for Fig. \ref{fig:crot_overp_ell.eps} but for the one-ring sequence. See also Tab. \ref{tab:cr}.}
\label{fig:crot_overp.eps}
\end{figure}

Figure \ref{fig:axis-ratio_he_n05.eps} displays the axis-ratio versus $\omega^2$ for the solutions discussed above. We clearly see that over-pressurization allows for equilibria beyond the classical limits as soon as $\hh_e >0$, as a direct consequence of the flattening of the mass density distribution. This is a major result. Table \ref{tab:cr} lists the axis ratios at critical rotation obtained for three values of external-to-core enthalpy (or pressure) ratio. Figure \ref{fig:crot_overp_ell.eps} shows the density structure for the states of critical rotation in the isolated case and for $\hh_e=0.25$. The polytropic index is $n=1.5$. The rings are shown in Fig. \ref{fig:crot_overp.eps}. When $\hp_e$ is large enough, the one ring-sequence is complete, meaning the possible existence of the intermediate hamburger sequence and its connection with ellipsoidal branch \citep{es81}. It would be interesting to determine, for each index $n$, the external-to-core enthalpy ratio required to make the path from ellipsoids to rings again continuous.

\begin{table}
  \caption{Axis ratio at critical rotation for the ellipsoid (E) and one-ring (R) sequences obtained for $n \in \{0.5,1.5\}$.}
  \centering
  \begin{tabular}{lllllll}
   &          &  & \multicolumn{2}{l}{\cite{hachisu86}} & \multicolumn{2}{l}{this work} \\
 $n$    &  $\hh_e$ &  $\hp_e$ &  E       &   R       & E       & R           \\\hline
 $0.5$  & $0$        &   $0$     &  $0.442$ &  $0.137$  & $0.437$ & $0.130$ \\
        & $0.25$     &   $0.125$ &         &           & $0.322$ & $0.022$ \\
        & $0.5$      &   $0.353$ &         &           & $0.268$ & $0$ \\
        & $0.75$     &  $0.650$ &         &            & $0.230$ & $0$  \\ \hline
 $1.5$  & $0$        &  $0$      & $0.617$  &  $0.325$  & $0.614$ & $0.325$ \\
        & $0.25$     &  $0.0312$ &          &           & $0.471$ & $0.214$ \\
        & $0.5$      &  $0.177$ &          &           & $0.368$ & $0.098$ \\
        & $0.75$     &  $0.487$ &          &           & $0.282$ & $0$  \\ \hline
  \end{tabular}
  \label{tab:cr}
\end{table}

\begin{figure}
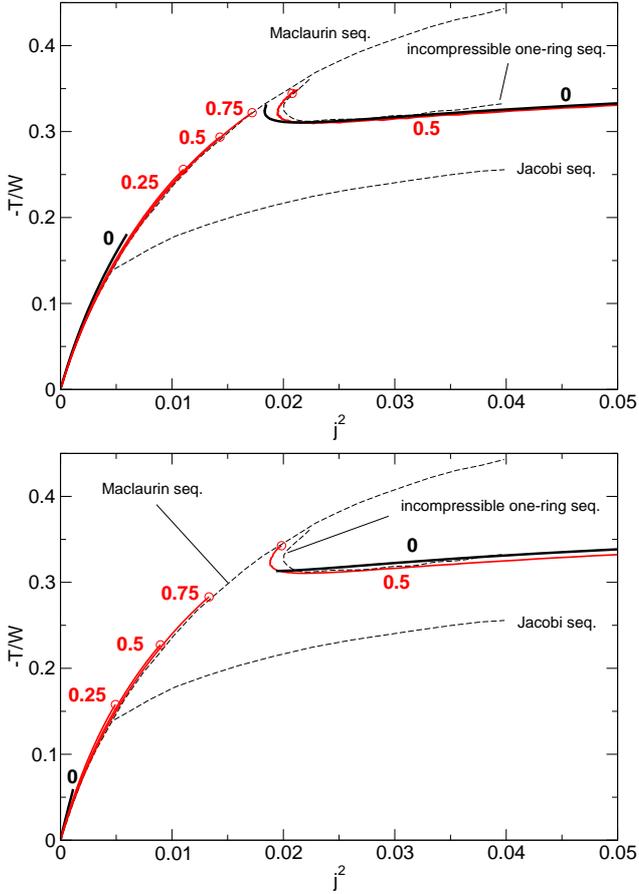

  \includegraphics[width=8.3cm,bb=33 34 726 530,clip=]{tw0.5.eps}\\
  \includegraphics[width=8.3cm,bb=33 34 726 530,clip=]{tw1.5.eps}
\caption{Stability indicator $T/|W|$ for a self-gravitating polytrope undergoing uniform external pressure at the fluid boundary ({\it red bold}) compared to the zero-pressure case ({\it black}). Values of $\hh_e$ are labelled on the curves. Rotation is uniform (i.e. solid). The polytropic index is $n=0.5$ ({\it top}) and $n=1.5$ ({\it bottom}). The incompressible Maclaurin, one-ring and Jacobi sequences are shown in comparison ({\it dashed lines}).}
\label{fig:tw.eps}
\end{figure}

  \subsection{Rough stability indicator}
  \label{sec:rsi}

Maclaurin ellipsoids are known to develop instabilities in a dynamical time when the kinetic-to-gravitational energy ratio $- T/W \equiv \beta \gtrsim 0.27$ \citep{chandra69}. This threshold is slightly reduced when general relativity effects are accounted for \citep{pa15,sy16}. With viscosity, systems are unstable for $\beta \gtrsim 0.14$, but this happens on very long time scales. Differential rotation allows for instabilities at much smaller values of the $\beta$-parameter, which is a recent discovery \citep{cen01,sy06,pa15,sy16}. The stability of rings is often investigated in presence of a central massive object \citep{shu83,sv91,chris93,ASNA:ASNA211}, to begin with Maxwell's problem for Saturn's rings \citep{vk07}. In \cite{th90}, polytropic rings with $n=1.5$ are found to be unstable for $\beta \gtrsim 0.16$.

The stability can be investigated numerically, by feeding a hydrodynamic code with equilibrium solutions \citep[e.g.][]{th90,pdd96}, or by analytical means which is more cumbersome. Clearly, both the equation of state and rotation profile play a key role. We will not handle this fundamental question in details. We limit ourselves to a simple plot of the $\beta$-parameter for the series of sequences discussed above. This is Fig. \ref{fig:tw.eps} (see also Tabs. \ref{tab:datae} and \ref{tab:datat}). We observe that $\beta(j)$ follows values for the incompressible branches, without significant deviations. This is especially pronounced for the ellipsoidal sequence. The new curves are more or less stretched along the Maclaurin branch depending on $\hh_e$, and $\beta$ clearly increases as $\hh_e$ rises. This is true whatever the polytropic index. In the limit $\hh_e \rightarrow 1$, over-pressurized fluids have the same stability properties as for the incompressible, zero-pressure analogues, while this case is mathematically singular (see Sect. \ref{subsec.pbatfb}). For low to moderate external stresses, no firm conclusion is, however, fairly possible. The analysis must actually be performed at fixed mass when perturbing $\hh_e$, and this quantity clearly varies along all sequence \citep{bonnor56,sta83b}. Conclusions are similar for the ring sequences, while a non-linearity is visible close to critical rotation.

\begin{table}
\centering
\begin{tabular}{llll}\\
  quantity                  & \cite{hachisu86}    & $\hh_e=0$   & $\hh_e=0.1$ \\ \hline
  covering factor $\Lambda$ & $0.091^*$          & $0.783$    & $0.788$  \\
  $C_1$                     & $0.505^*$            & $0.5037$   & $0.6207$   \\
  $C_2$ (i.e. $\Omega_0^2L^2)$  & $0.215$           & $0.2154$ & $0.2475$   \\
  $-C_3$                     & ?                 & $0.6740$   & $0.8212$    \\
  $\hr_e$                    &  $1$              & $1$ & $1$\\
  $\hr_p$                    &  $\frac{1}{3}$    & $\frac{1}{3}$ &$\frac{1}{3}$\\
 $\hbsur$                   &  ?                   & $0.8707$ & $0.8757$\\
  $\hv$                      & $2.61$             & $2.6089$ & $2.6248$ \\
  $\hm$                      & $0.639$             & $0.6391$ & $0.8312$\\
  $\langle \hrho \rangle$    & $0.245^*$           & $0.2450$ & $0.3167$\\
  max. pressure              & $0.202$             & $0.2015$ &$  0.2483$\\
  max. density               & $0.359^*$           & $0.3574$ & $  0.4890$\\
  $\sqrt{C_2}\hjcin$         & $0.137$             & $0.1373$ & $  0.2067$\\
  $C_2 \hcin$                & $0.0638$            & $0.0638$ & $  0.0962$\\
  $-\hw$                     & $0.372$             & $0.3716$ & $  0.5800$\\
 $\beta$                    & $0.17$              & $0.172$  &  $0.166$ \\
  $\frac{C_1}{n+1}\hu$       & $0.244$             & $0.2440$ & $  0.3939$\\
  $-\Pi_e$                   & $0$                  &$0$        & $ 0.0062$\\
  $\log(VP)$                 & ?                  & $-4.44$ &   $ -4.61$\\
 iterations & ?              & $44$   &  $ 35$\\\hline
 $^*$estimated.\\
\end{tabular}
\caption{Results for the ellipsoids shown in Fig. \ref{fig:config_overp_ell_v.eps} corresponding to the $v$-constant rotation law. The polytropic index is $n=1.5$. The last column is for the over-pressurized fluid.}
\label{tab:datae-v}
\end{table}

\begin{figure*}
  \centering
  \includegraphics[height=5.1cm,bb=76 265 483 675, clip==]{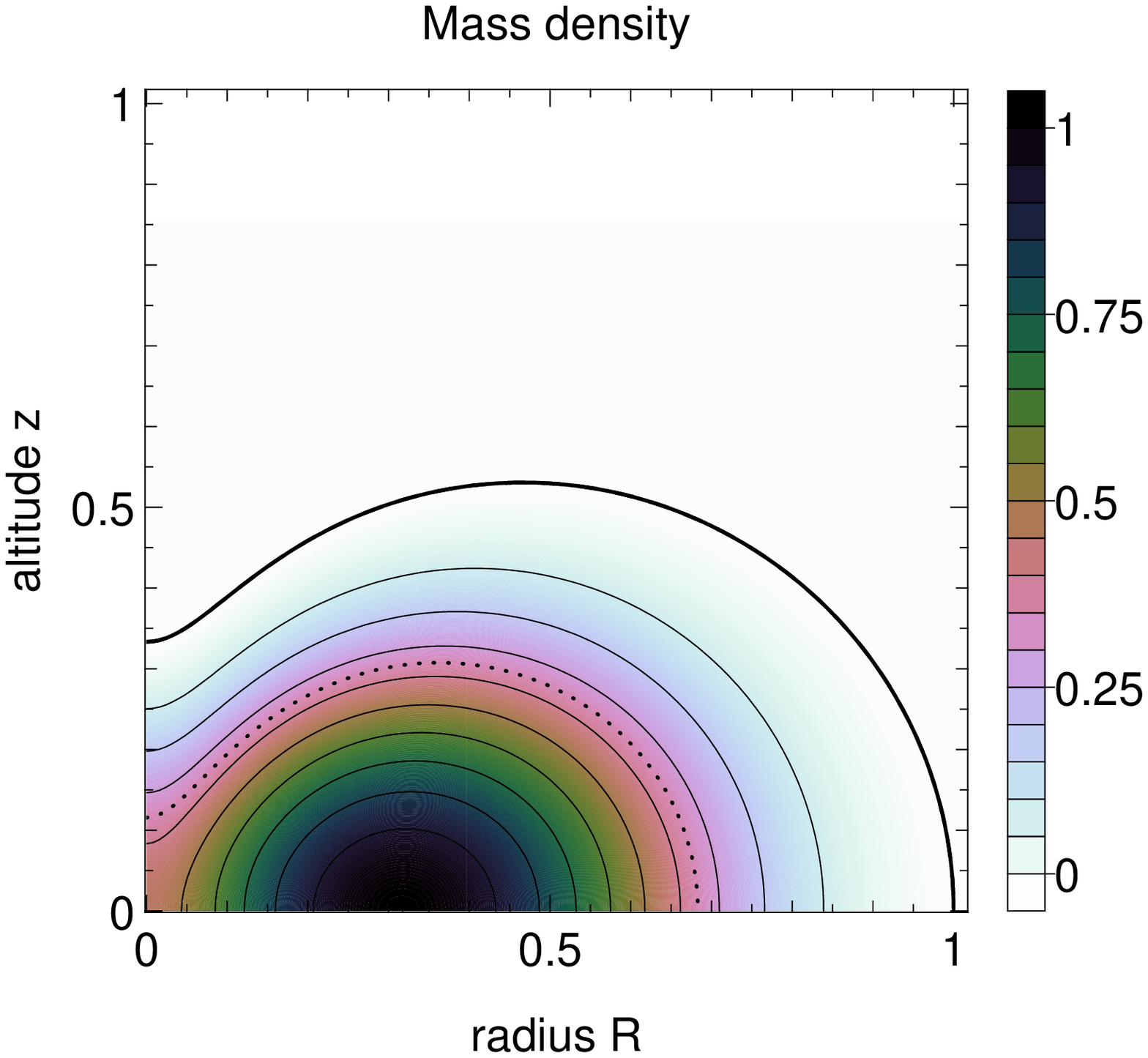}\hspace*{25pt}\includegraphics[height=5.1cm,bb=76 265 483 675, clip==]{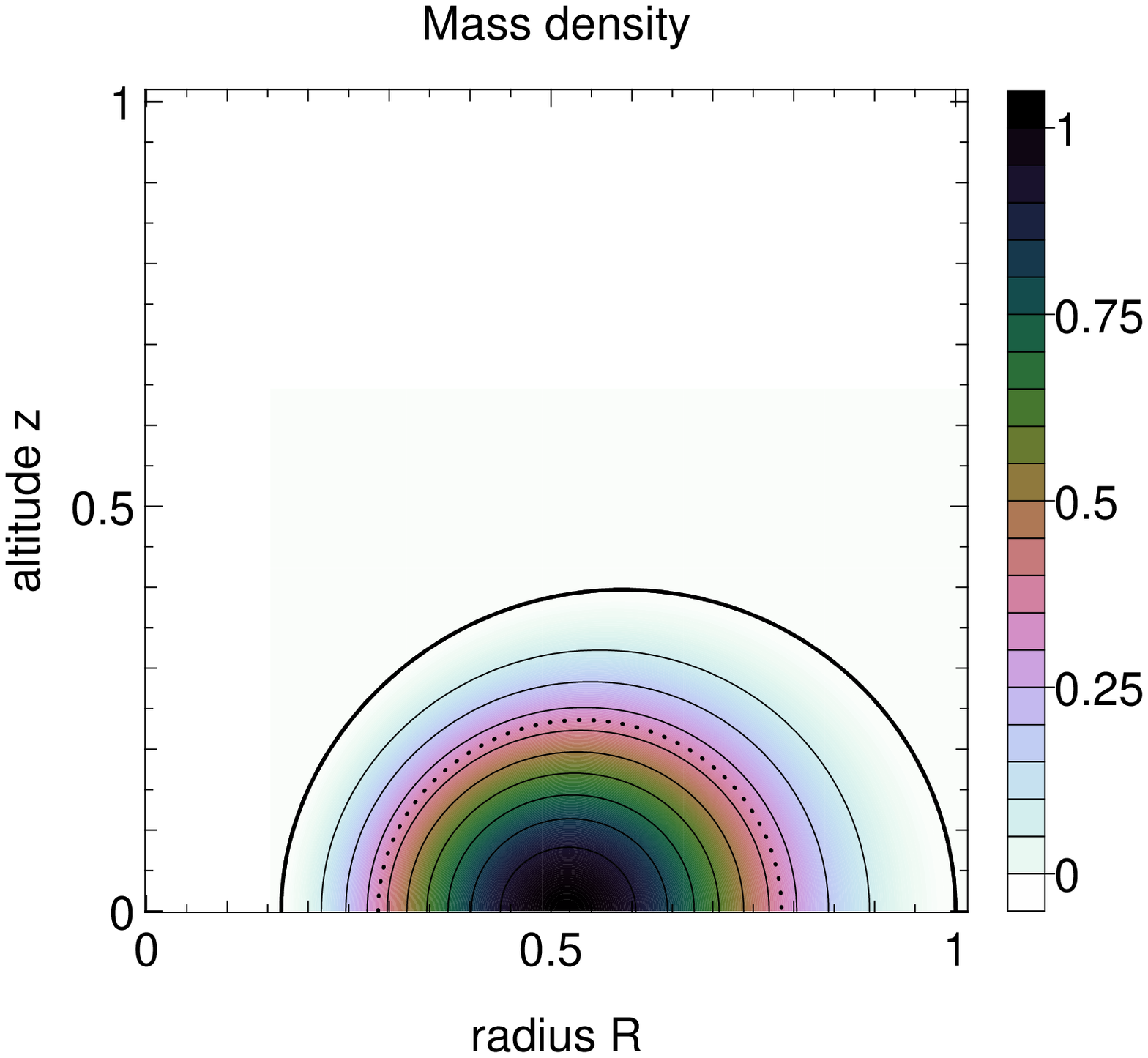}\\
  \bigskip
  \hspace*{25pt}\includegraphics[height=5.1cm,bb=76 265 553 675,clip=true]{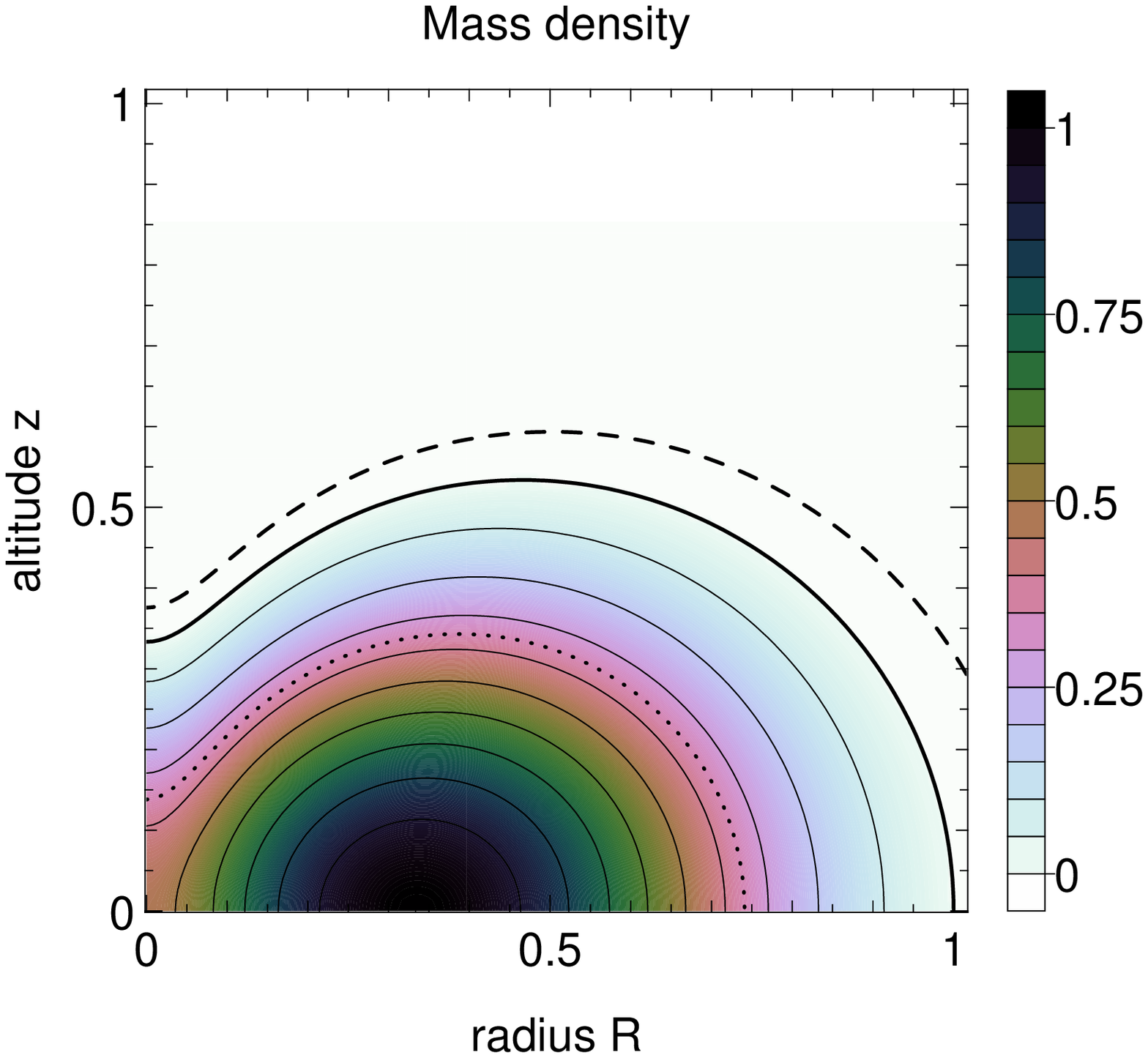}\includegraphics[height=5.1cm,bb=76 265 553 675,clip=true]{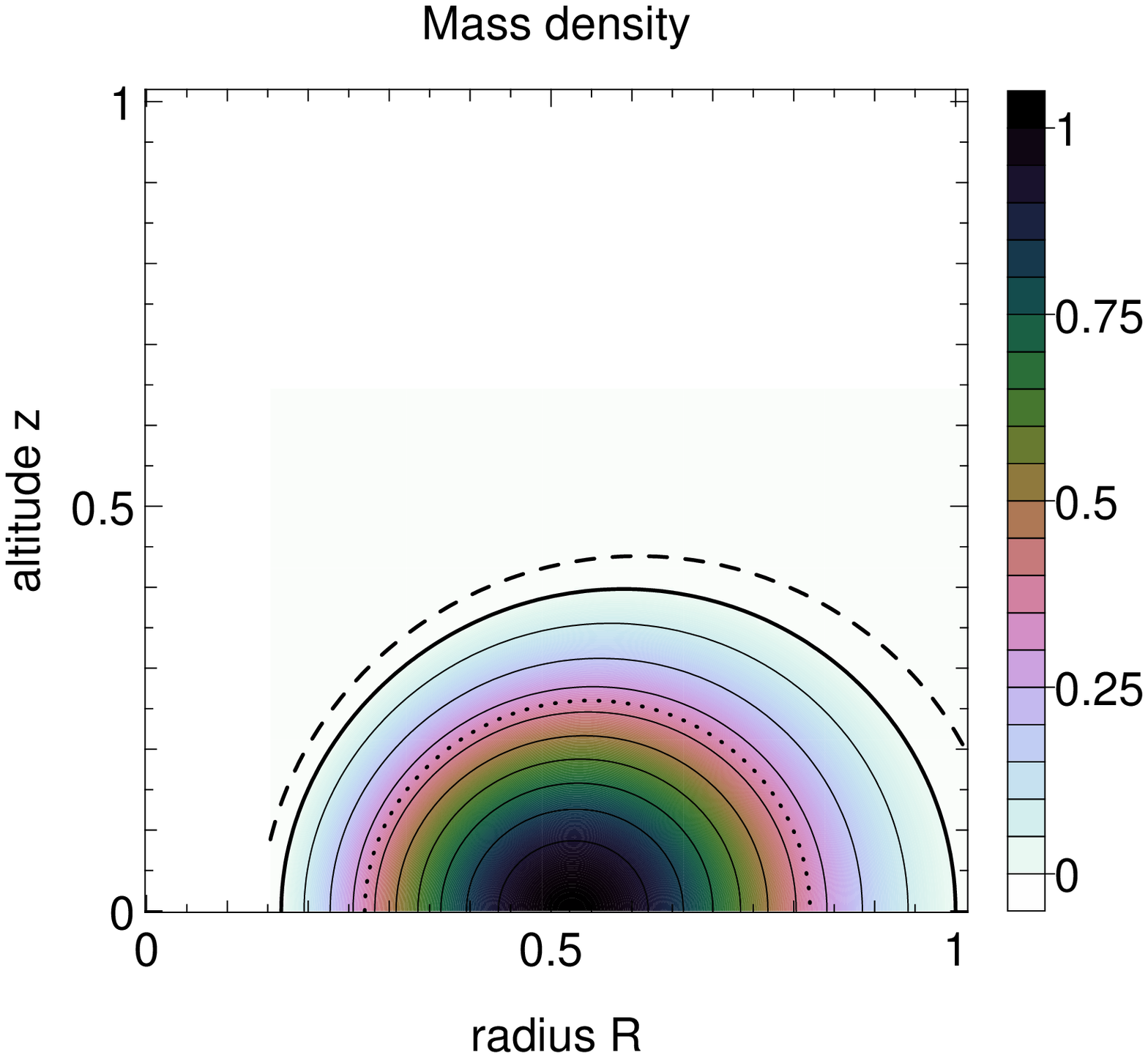}
\caption{Mass density structure for the ellipsoid ({\it left}) and for the ring ({\it right}) rotating following the $v$-constant rotation law in the isolated case ({\it top}) and with uniform external pressure corresponding to $\hh_e=0.1$ ({\it bottom}). The polytropic index is $n=1.5$, the axis ratio is $\hz_A/\hr_B=\frac{1}{3}$ for the ellipsoid, $\hr_A/\hr_B=\frac{1}{6}$ for the ring, and the softening parameter is $\hd=0.1$. Density contours are every $\Delta \hrho=0.1$ ({\it thin lines}). Also shown are the fluid boundary where $\hh - \hh_e=0$ ({\it bold}), the zero enthalpy level ({\it dashed}), and the place where $\hh=0.5$ ({\it dotted}), which corresponds to $\hrho \approx 0.353$. See also Tabs. \ref{tab:datae-v} and \ref{tab:datat-v}.}
\label{fig:config_overp_ell_v.eps}
\end{figure*}

\begin{table}
\centering
\begin{tabular}{llll}\\
  quantity                  & \cite{hachisu86}     & $\hh_e=0$   & $\hh_e=0.1$ \\ \hline
  covering factor $\Lambda$ & $0.091^*$          & $0.699$    & $0.701$  \\
  $C_1$                     & $0.297^*$            & $0.2966$   & $0.3609$   \\
  $C_2$ (i.e. $\Omega_0^2L^2)$  & $0.233$         & $0.2333$   & $0.2659$   \\
  $-C_3$                     & ?                 & $0.5934$   & $0.6980$    \\
  $\hr_e$                    &  $1$              & $1$ & $1$\\
  $\hr_p$                    &  ?    & $0.397$ &$0.398$\\
 $\hbsur$                   &  ?                 & $0.5176$ & $0.5188$\\
  $\hv$                      & $1.90$            & $1.9030$ & $1.9079$ \\
  $\hm$                      & $0.538$            & $0.5380$ & $0.6616$\\
  $\langle \hrho \rangle$    & $0.283^*$          & $0.2827$ & $0.3468$\\
  max. pressure              & $0.119$            & $0.1186$ &$  0.1444$\\
  max. density               & $0.162^*$          & $0.1615$ & $  0.2168$\\
  $\sqrt{C_2}\hjcin$         & $0.150$            & $0.1497$ & $  0.2024$\\
  $C_2 \hcin$                & $0.0606$           & $0.0606$ & $  0.0850$\\
  $-\hw$                     & $0.246$            & $0.2464$ & $  0.3558$\\
  $\beta$                    & $0.25$             & $0.246$  & $0.239$ \\
  $\frac{C_1}{n+1}\hu$       & $0.125$            & $0.1252$ & $  0.1884$\\
  $-\Pi_e$                   & $0$                &$0$        & $ 0.0026$\\
  $\log(VP)$                 & ?                  & $-4.40$ &   $ -4.00$\\
 iterations & ?              & $44$   &  $ 35$\\\hline
 $^*$estimated.\\
\end{tabular}
\caption{Results for the rings shown in Fig. \ref{fig:config_overp_ell_v.eps} corresponding to the $v$-constant rotation law. The polytropic index is $n=1.5$.  The last column is for the over-pressurized fluid.}
\label{tab:datat-v}
\end{table}

\begin{table}
\centering
\begin{tabular}{llll}\\
  quantity                  & \cite{hachisu86}     & $\hh_e=0$   & $\hh_e=0.1$ \\ \hline
  covering factor $\Lambda$ & $0.091^*$          & $0.735$    & $0.739$  \\
  $C_1$                     & $0.847^*$            & $0.8465$   & $1.0494$   \\
  $C_2$ (i.e. $\Omega_0^2L^2)$  & $0.0145$        & $0.0145$ & $0.01672$   \\
  $-C_3$                     & ?                 & $0.8491$   & $1.0047$    \\
  $\hr_e$                    &  $1$              & $1$ & $1$\\
  $\hr_p$                    &  ?    & $0.815$ & $0.818$ \\
 $\hbsur$                   &  ?                   & $1.2747$ & $1.2808$\\
  $\hv$                      & $3.74$             & $3.7395$ & $3.7488$ \\
  $\hm$                      & $0.838$             & $0.8384$ & $1.0494$\\
  $\langle \hrho \rangle$    & $0.224^*$           & $0.2242$ & $0.2919$\\
  max. pressure              & $0.339$             & $0.3386$ &$  0.4198$\\
  max. density               & $0.780^*$           & $0.7788$ & $  1.0750$\\
  $\sqrt{C_2}\hjcin$         & $0.0928$             & $0.09283$ & $  0.1313$\\
  $C_2 \hcin$                & $0.0391$            & $0.03910$ & $  0.0537$\\
  $-\hw$                     & $0.596$             & $0.59550$ & $  0.9326$\\
  $\beta$                    & $0.066$             & $0.0657$ &  $0.0576$\\
  $\frac{C_1}{n+1}\hu$       & $0.517$             & $0.51732$ & $  0.8401$\\
  $-\Pi_e$                   & $0$                  &$0$        & $ 0.0149$\\
  $\log(VP)$                 & ?                  & $-4.35$ &   $ -4.21$\\
 iterations                  & ?              & $31$   &  $ 32$\\\hline
  $^*$estimated.\\
\end{tabular}
\caption{Results for the ellipsoids shown in Fig. \ref{fig:config_overp_ell_j.eps}  and corresponding to the $j$-constant rotation law. The polytropic index is $n=1.5$. The last column is for the over-pressurized fluid.}
\label{tab:datae-j}
\end{table}

\begin{figure*}
  \centering
  \includegraphics[height=5.1cm,bb=76 265 483 675, clip==]{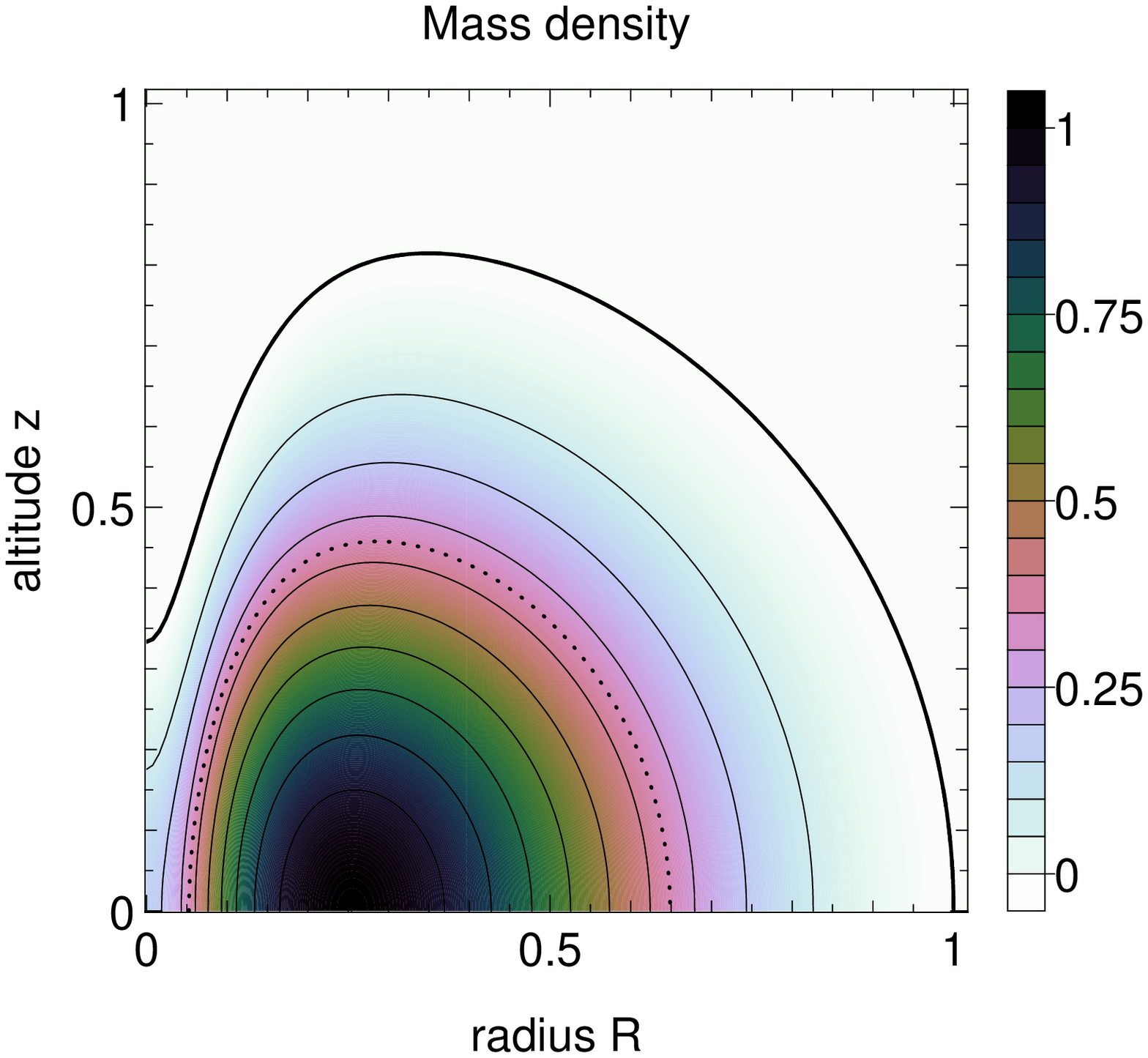}\hspace*{25pt}\includegraphics[height=5.1cm,bb=76 265 483 675, clip==]{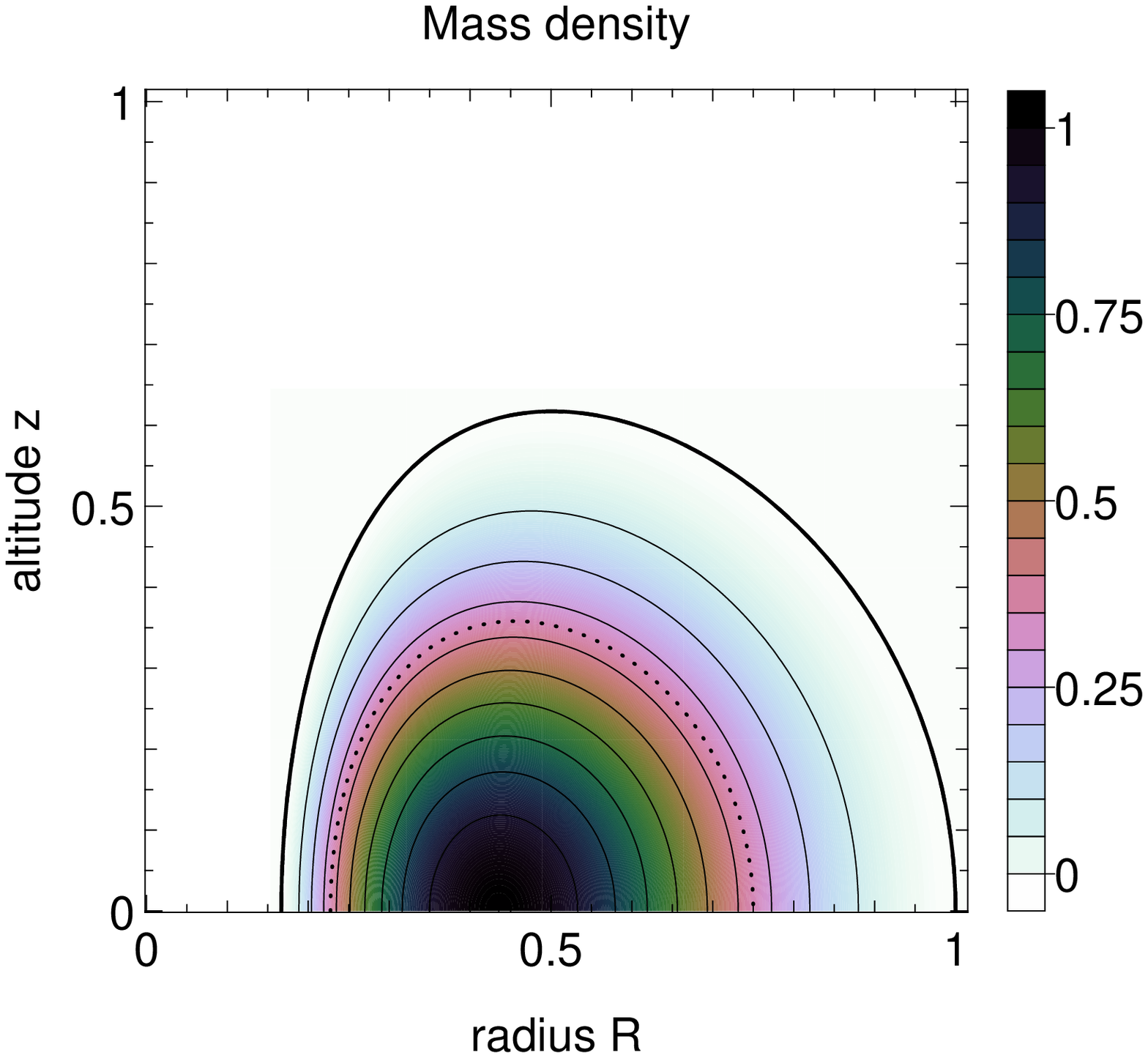}\\
  \bigskip
  \hspace*{25pt}\includegraphics[height=5.1cm,bb=76 265 553 675,clip=true]{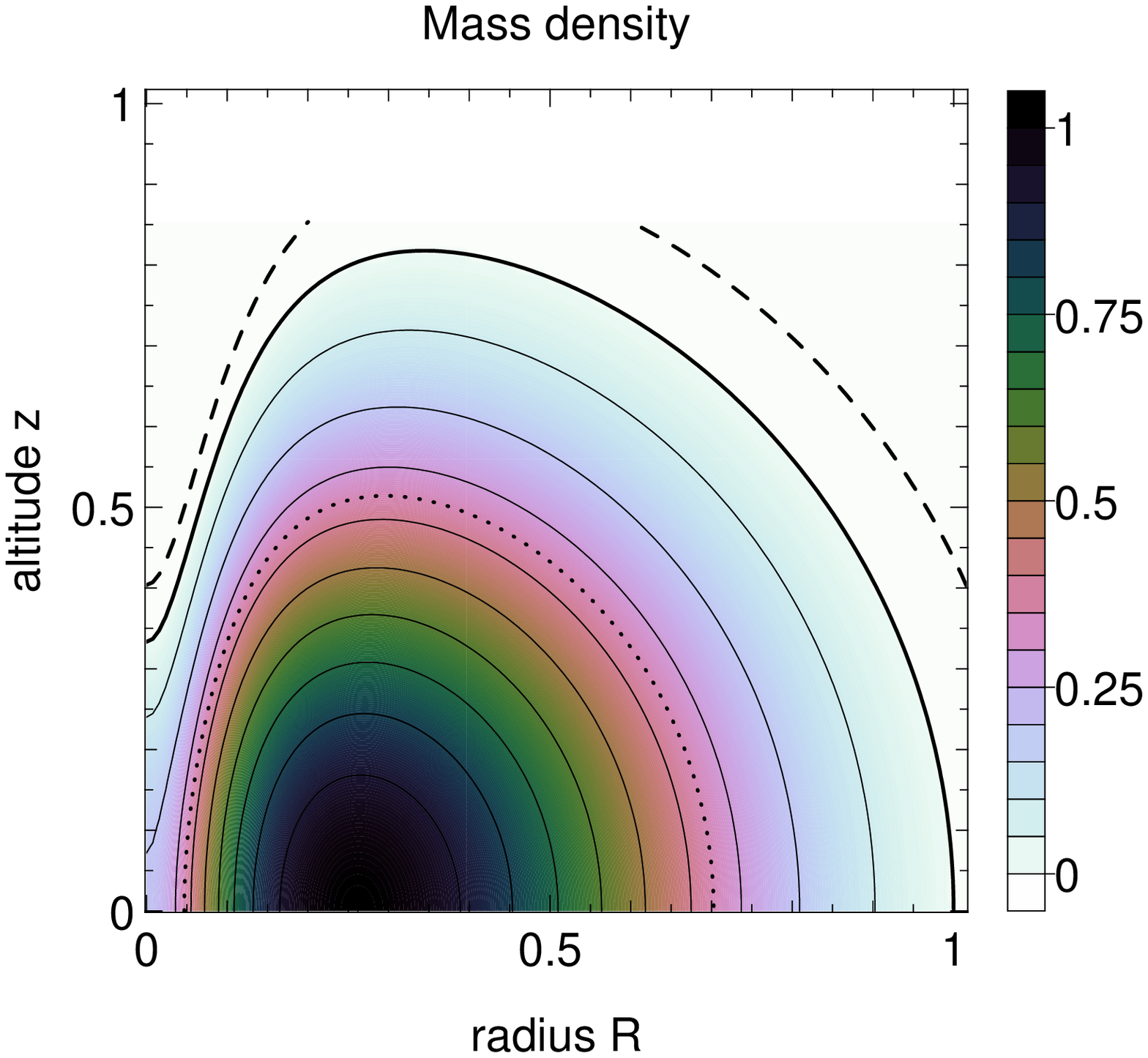}\includegraphics[height=5.1cm,bb=76 265 553 675,clip=true]{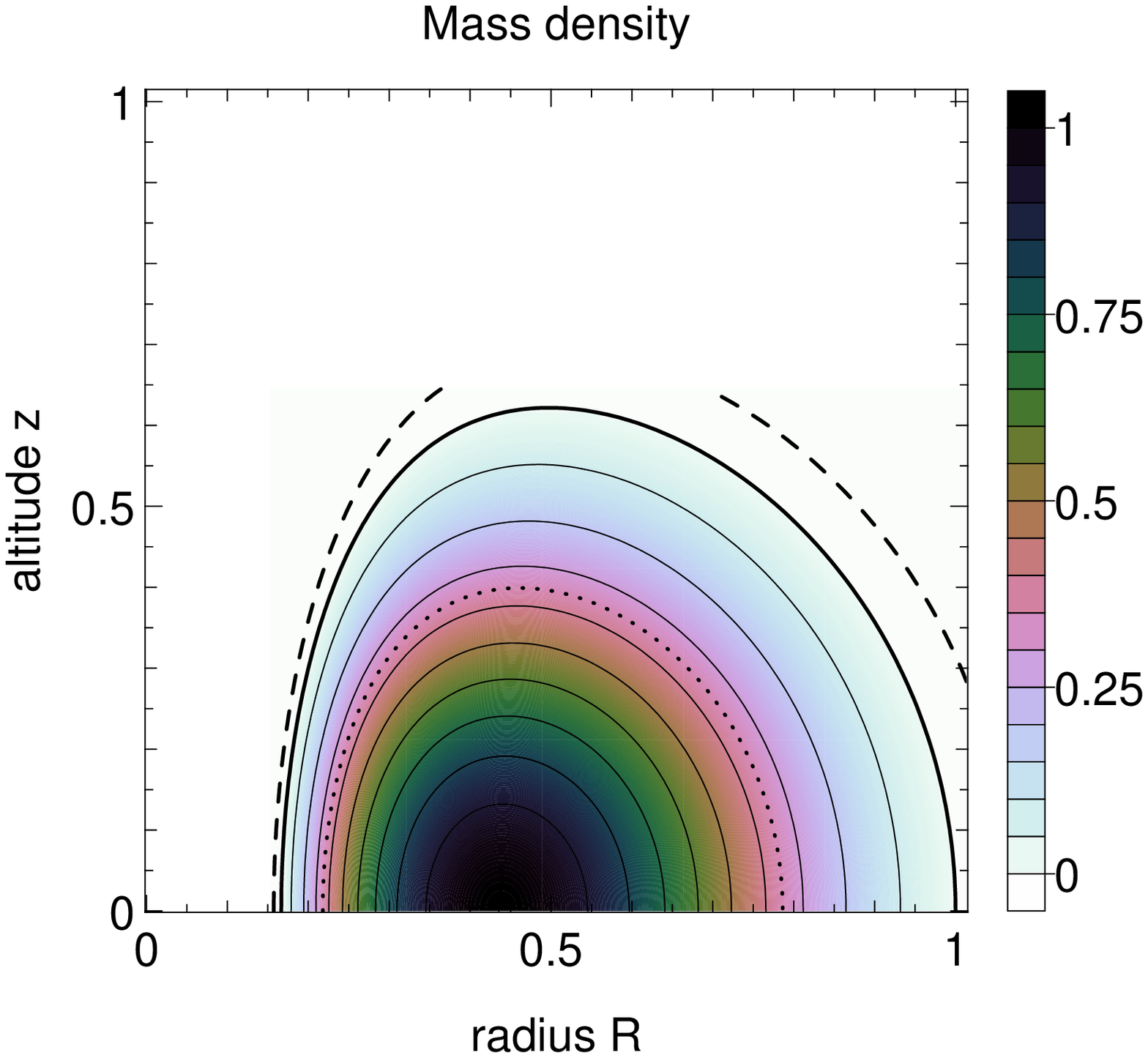}
\caption{Same legend as for Fig. \ref{fig:config_overp_ell_v.eps} but for the  $j$-constant rotation law. See also Tabs. \ref{tab:datae-j} and \ref{tab:datat-j}.}
\label{fig:config_overp_ell_j.eps}
\end{figure*}

\begin{table}
\centering
\begin{tabular}{llll}\\
  quantity                  & \cite{hachisu86}     & $\hh_e=0$   & $\hh_e=0.1$ \\ \hline
  covering factor $\Lambda$ & $0.091^*$          & $0.728$    & $0.734$  \\
  $C_1$                     & $0.565^*$          & $0.5661$   & $0.6946$   \\
  $C_2$ (i.e. $\Omega_0^2L^2)$  & $0.0504$         & $0.0504$   & $0.0567$   \\
  $-C_3$                     & ?                 & $0.8038$   & $0.9291$    \\
  $\hr_e$                    &  $1$              & $1$       & $1$\\
  $\hr_p$                    &  ?                & $0.617$  &$0.622$\\
  $\hbsur$                   &  ?                & $0.8083$ & $0.8151$\\
  $\hv$                      & $2.87$            & $2.8677$  & $2.8859$ \\
  $\hm$                      & $0.780$            & $0.7798$ & $0.9652$\\
  $\langle \hrho \rangle$    & $0.271^*$          & $0.2719$ & $0.3344$\\
  max. pressure              & $0.226$            & $0.2264$ & $0.2779$\\
  max. density               & $0.425^*$          & $0.4260$ & $0.5789$\\
  $\sqrt{C_2}\hjcin$         & $0.168$            & $0.1677$ & $0.2205$\\
  $C_2 \hcin$                & $0.0775$           & $0.7751$ & $0.1036$\\
  $-\hw$                     & $0.497$            & $0.4973$ & $0.7207$\\
  $\beta$                    & $0.16$             & $0.1558$ & $0.144$ \\
  $\frac{C_1}{n+1}\hu$       & $0.342$            & $0.3423$ & $  0.5211$\\
  $-\Pi_e$                   & $0$                &$0$        & $ 0.0076$\\
  $\log(VP)$                 & ?                  & $-4.40$ &   $ -3.97$\\
  iterations & ?              & $35$   &  $ 34$\\\hline
  $^*$estimated.\\
\end{tabular}
\caption{Results for the ring shown in Fig. \ref{fig:config_overp_ell_j.eps} corresponding to the $v$-constant rotation law. The last column is for the over-pressurized fluid.}
\label{tab:datat-j}
\end{table}

\section{Differential rotation}
\label{sec:difrot}

A defect in the present problem is the ad-hoc character of the rotation law which is hard to prescribe. The dynamics is in principle regulated by the fluid itself and depends on the past history of the system (initial conditions and evolution) that is not known. This situation, somewhat similar for galaxies, is typical of self-gravitating systems. Here, we go beyond the case of rigid rotation and consider two among the most employed rotation laws : the $v$-constant and the $j$-constant laws. Other options are possible \citep{sta83a}.

For a constant azimuthal velocity, $\Omega \propto 1/R$. Due to the singularity onto the $z$-axis, this profile needs to be softened for ellipsoidal configurations. The rotation rate is generally defined by $\Omega \sqrt{R^2+d^2} =cst$ where $d$ is a free parameter. The associated centrifugal potential, in dimensionless form, is 
\begin{equation}
  \hphi= - \ln \sqrt{\hr^2+\hd^2},
  \label{eq:phiv}
\end{equation}
where $L \hd =d$. For a uniform specific angular momentum in the fluid, we have $\Omega \propto 1/R^2$. Softening is also necessary for the same reasons. According to \cite{hachisu86}, we take:
\begin{equation}
  \hphi= \frac{1}{2(\hr^2+\hd^2)}.
  \label{eq:phij}
\end{equation}

We have run simulations for many axis ratios and polytropic indices by using Eqs.(\ref{eq:phiv}) and (\ref{eq:phij}). We present two configurations reported in \cite{hachisu86}. For the $v$-constant profile, we select the ellipsoid with $\hz_A/\hr_B=\frac{1}{3}$ and the ring with $\hr_A/\hr_B=\frac{1}{6}$, and the parameter is $\hd=0.1$ in both cases. The results are summarised in Tabs \ref{tab:datae-v} and \ref{tab:datat-v} for $\hh_e \in \{0,0.1\}$. The mass density distributions are plotted in Fig. \ref{fig:config_overp_ell_v.eps}. Regarding the $j-$constant profile, the conditions are the same. Output quantities are given in Tabs. \ref{tab:datae-j} and \ref{tab:datat-j}. The internal structures are displayed in Fig. \ref{fig:config_overp_ell_j.eps}. The effects of over-pressurization are globally similar as in the case of rigid rotation. The fluid is more massive, rotates faster and is slightly thicker due to broader potential wings. The mass shedding limit is pushed back. For instance, for the $v$-constant profile, the equilibrium sequences are closed for $n \lesssim 2.68$ in the absence of external pressure, and one goes continuously from the ellipsoidal to the one ring sequence. For $\hh_e=0.1$, the sequence breaking occurs still inside the ellipsoid part of the $(\omega^2,j^2)-$diagram, but at $n \approx 3.24$. The $\beta$-parameter is lower. This does not mean that systems are more stable since masses are not conserved (see Sect. \ref{sec:rsi}).

\section{Typical reaction in presence of a photon point source}
\label{sec:ips}

\begin{figure}
\includegraphics[width=8.4cm,bb=0 0 475 347,clip=]{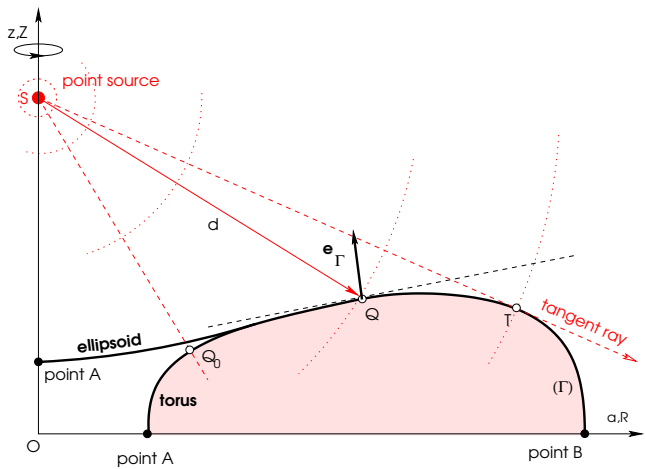}
\caption{Configuration for the polytrope over-pressurized by photon isotropically emitted by a point source S. Beyond point T of the boundary where $\vec{\rm SQ}.\vec{e}_\Gamma=0$, the system does not see the source and is not impacted by photons. At point Q$_0$, the separation between the boundary and the source is the shortest with $\vec{\rm SQ}.\vec{e}_\Gamma=-1$ and the external enthalpy is at the boundary is denoted $\hh_e^0$.}
\label{fig:point_sources.eps}
\end{figure}

We consider a point source in the vicinity of the self-gravitating polytrope. Such a situation is common. During star formation as well as in more evolved systems like in binaries, disks, tori and rings are inevitably irradiated \citep{smak89,betal13}. A light source of non-stellar origin is present along the rotation axis, like jets, shocks, or a X-ray lamp-post in the AGN context \citep{rozanska02,goosman06}. Even in the assumption of a perfectly reflecting surface, the problem is not easy to solve self-consistently here. Not only $\hh_e$ all along $(\Gamma)$ depends on the distance from the source but the inclination of rays with respect to the local normal of the boundary is also involved. This is the kind of situation where numerical instabilities can occur \citep[e.g.][]{dull00}. The underlying SCF-method is possibly not the nominal way to handle this question. Since our purpose is purely illustrative, we restrict the discussion to the zero-order effect and proceed in two steps. First, we determined the equilibrium of the isolated fluid (i.e. without a point source) and then determine the semi-major axis $a$, semi-minor axis $b$ and centre $C(R_c,Z_c)$ of the best ellipse that fits the boundary $(\Gamma$). Obviously, this has sense far enough from critical rotations where shapes are indeed elliptical. Then, we calculate the unit vector $\vec{e}_\Gamma$, normal to ($\Gamma)$ and oriented outward. If Q$(\vec{r}_\Gamma)$ is a point of the boundary and $\vec{r}_S(R_S,Z_S)$ denotes the position of the source (a point or a loop under axial symmetry), then the general expression for the cosine is
\begin{equation}
\cos(\vec{\rm SQ},\vec{e}_\Gamma) = \frac{(R-R_S)(R-R_c)b^2+(Z-Z_S)(Z-Z_c)a^2}{SQ \sqrt{a^4(Z-Z_c)^2+b^4(R-R_c)^2}}.
\end{equation}
 For this point source model, we take
\begin{equation}
\frac{\hp_e(\hr,\hz)}{\hp_e^0} = \frac{1}{d^2} \left\{ \sup \left[0,-\cos(\vec{SQ},\vec{e}_\Gamma)\right] \right\}^2,
\label{eq:he_illps}
\end{equation}
where $d=SQ/SQ_0 \ge 1$ and $SQ_0$ is the shortest distance. With this formula, the flux of photons is conserved on concentric spheres centred on the actual point source S, and $\hp_e(\hr,\hz) =  \hp_e^0$ at point $Q_0$ where $\vec{SQ}$ and $\vec{e}_\Gamma$ are co-linear. Besides, only parts of the fluid surface that directly see the source are influenced. The situation is depicted in Fig. \ref{fig:point_sources.eps}. To conform to equatorial symmetry, there are in fact two sources here, the one at $(\hr_S,\hz_S)$ and the other at $(\hr_S,-\hz_S)$.

\begin{figure}
 \centering
 \includegraphics[height=5.1cm,bb=76 265 483 675, clip==]{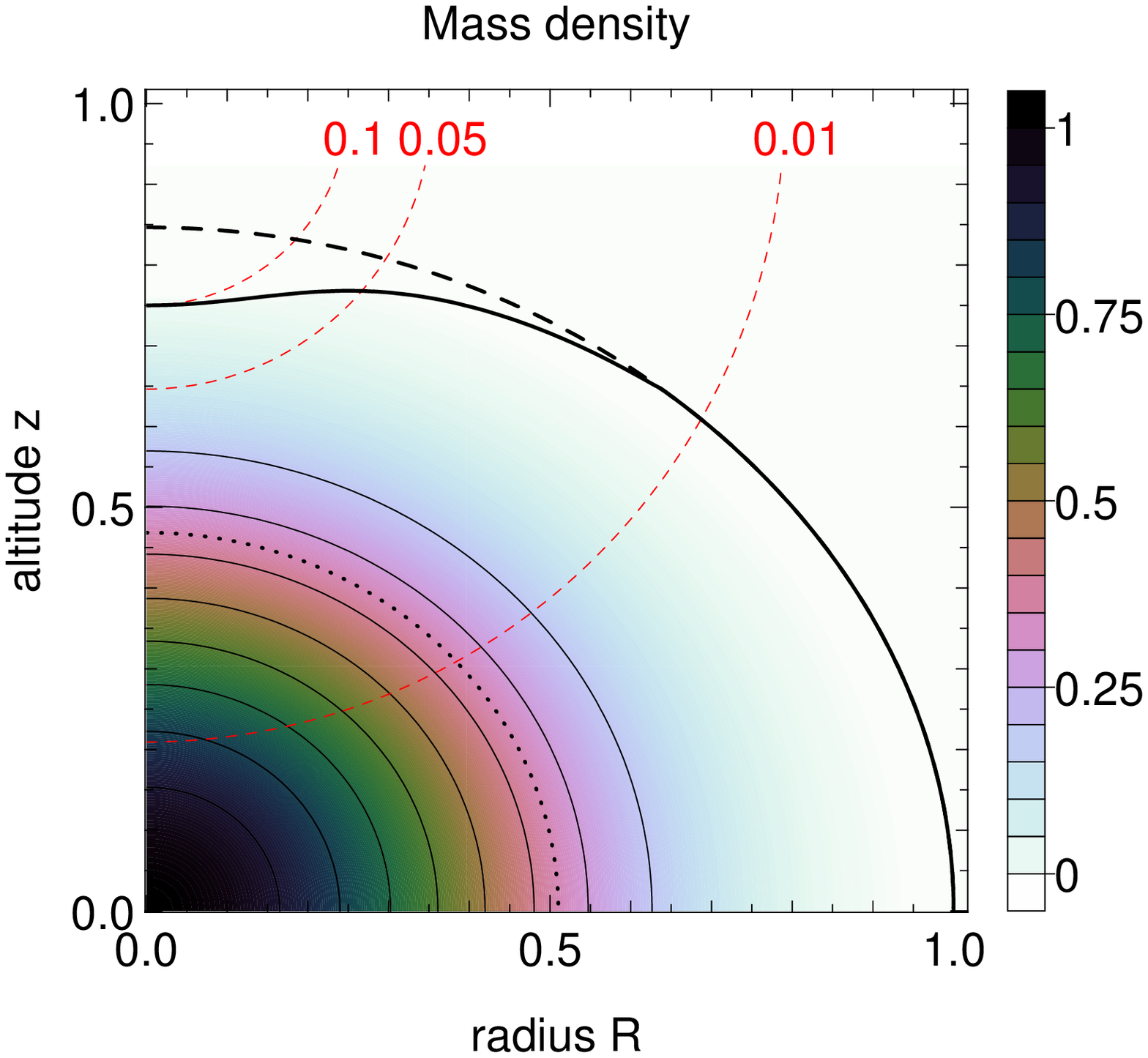}\\
 \bigskip
 \hspace*{25pt}\includegraphics[height=5.1cm,bb=76 265 553 675,clip=true]{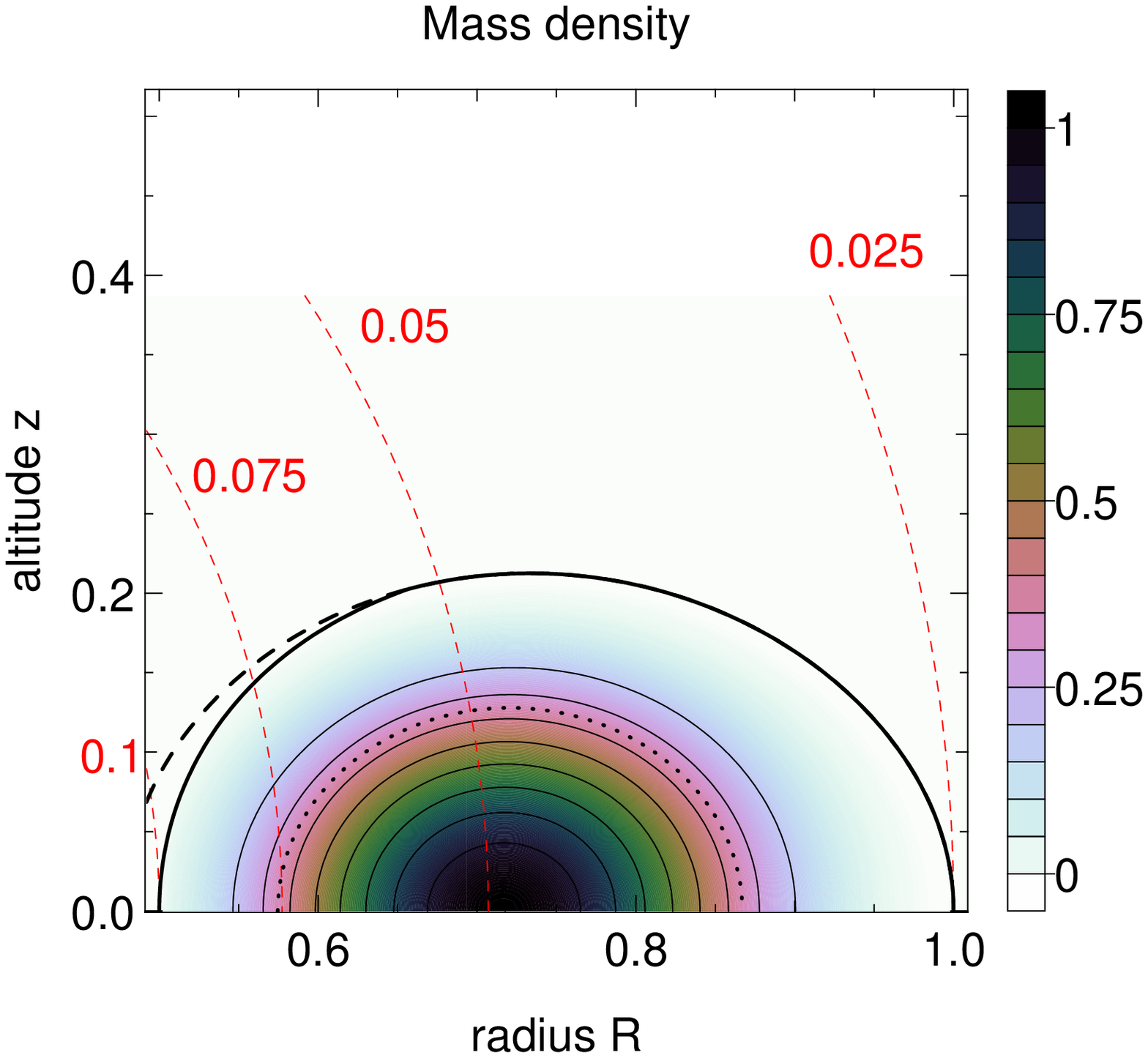}
\caption{Mass density structure for an ellipsoid ({\it top}) and a ring ({\it bottom}) over-pressurized by photons emitted by a point source. Rotation is rigid and the polytropic index is $n=1.5$. For the ellipsoid, the axis-ratio is $0.75$, the sources are located at $(0,\pm 1)$. For the ring, the axis ratio is $0.5$, the source stands at the origin. We take $\hh_e^0=0.1$ in both cases. Density contours are every $\Delta \hrho=0.1$ ({\it thin lines}). Also shown are the fluid boundary where $\hh - \hh_e=0$ ({\it bold}), the zero enthalpy level ({\it dashed}), the place where $\hh=0.5$ ({\it dotted}), which corresponds to $\hrho \approx 0.353$, and lines of constant external enthalpy $\hh_e$ ({\it red dotted}).}
\label{fig:rho_pointsource_both.ps}
\end{figure}

We have run the code for the two reference configurations considered in Sect. \ref{subsec:ellipsoid} and \ref{subsec:torus}, when external pressure prescribed by Eq.(\ref{eq:he_illps}). The enthalpy field $\hh_e$ is deduced by using  Eq.(\ref{eq:ph}). Rotation is rigid. For the ellipsoid, the dimensionless coordinates of the source are $(0,\pm 1)$. For the ring, the source is at the origin. We take $\hh_e^0=0.1$ in both cases. Figure \ref{fig:rho_pointsource_both.ps} show the resulting mass density maps. We clearly distinguish the region illuminated by the source from the one in the shadow. However, all the fluid participates in the new equilibrium, even the shadowed region, through the gravitational potential which is global. A quick look at the output quantities indicates that the properties of the ellipsoid are in between the isolated case and the case with uniform pressure corresponding to $\hh_e^0$. There is a jump in the density when crossing the fluid surface at the pole. Then $\hrho(\Gamma)$ gradually falls to zero from the pole to the tangent rays down. When $\hh_e^0$ is increased from $0$, the polar region gets flatter and flatter, and even have concave shape, like in the figure. We observe the same phenomenon for the toroidal configuration, the inner edge is sharper and shaper, as it is detected in some circumstellar disks.

\section{Conclusion}

This paper is devoted to self-gravitating polytropes embedded in an ambient pressure field. The external stress is assumed to be due to photons that, in contrast to gas, do not modify the gravitational field. Except symmetry considerations and the rigid rotation law, the main physical assumption concerns the albedo of the fluid surface: photons just hit the fluid surface an deposit momentum without penetrating the system. As for spherical systems, the condition of pressure balance at the boundary defines its location. Because of rotation, however, the equation of this boundary cannot be known in advance and needs a specific computation. As shown, equilibrium structures can be determined from the Bernoulli equation coupled to the Poisson equation. The SCF-method, in its classical form at least, is unable to solve the present equation set for any stress, as explained. In this purpose, we have defined a re-scaling operator to be inserted in the main SCF-loop. It acts like Dirichlet conditions by holding fixed the enthalpy field at a few selected space points. A more general SCF-method has been proposed. On this ground, we have successfully computed equilibrium structures for both ellipsoidal and toroidal/ring configurations. The main conclusion is that, in the dimensionless space, over-pressurization make the fluids more massive and with enhanced rotation rate, for a given axis ratio. We have also calculated the position of these over-pressurized solutions in the $\omega^2-j^2$ diagram. As shown, each new sequence is located in between the incompressible branch and the compressible, zero-zero-pressure branch. A major result is that states of critical rotations are exceeded as soon $P_e >0$.

This work can certainly be expanded in some directions. A first program would be to perform a stability analysis, especially for ring configurations which is rarely done. Following the standard approach, this question can be touched by constructing precise external pressure-volume diagrams, for a given total mass \citep[e.g.][]{mcc57}. Between two infinitely close over-pressurized equilibria, the total angular momentum and rotation rate is expected to change \citep[see e.g.][]{bo73}. This means an extra assumption. Another option would be to inject the solution into an hydrodynmic code \citep{th90, pdd96}. Regarding $\omega^2-j^2$ diagrams, it would be interesting to determine which minimum values of the control parameter (the external-to-core pressure ratio) are required to make the sequence closed again, as these are in the incompressible case. Another interesting perspective concerns photons. We have assumed that these deposit momentum onto the boundary layer without going through it. A noticeable improvement would be to relax this hypothesis. It is surely possible to build a double-layer model including a core and photosphere with specific polytropic index \citep[e.g.][]{kiu10}. In general, the appropriate value for a radiation pressure dominated medium is $n=3$ \citep{cox1968}. The extended SCF-method presented here is a priori suited to this kind of problem. Finally, regarding the point source model, it is necessary to go beyond the assumption made in the last section. The equation of the actual boundary is not that of an ellipse, in particular close to critical rotation. The precise shape must be acccounted for in the determination of the flux of photons through the cosine. There are interesting technical aspects.

\section*{Acknowledgements}
We would like to dedicate this paper to our friend J.P. Zahn. We thank the anonymous referee for his motivating and constructive report leading to significant improvements.

\bibliographystyle{aa}


\end{document}